\documentclass[]{interact}

\usepackage[caption=false]{subfig}

\usepackage[numbers,sort&compress]{natbib}
\bibpunct[, ]{[}{]}{,}{n}{,}{,}
\makeatletter
\def\NAT@def@citea{\def\@citea{\NAT@separator}}
\makeatother

\theoremstyle{plain}

\theoremstyle{definition}

\theoremstyle{remark}

\usepackage{enumerate}

\usepackage{amsmath}
\usepackage{amssymb}

\usepackage{cancel}
 \usepackage{caption}
\usepackage{ifpdf}
\usepackage{rotating}
\usepackage{alltt}
\usepackage{units}
\usepackage{color}
\usepackage[usenames,dvipsnames]{xcolor}

\definecolor{gray}{rgb}{0.3, 0.3, 0.3}
\definecolor{lgray}{gray}{0.7}
\definecolor{llgray}{gray}{0.5}
\definecolor{lllgray}{gray}{0.3}

\definecolor{c1}{rgb}{0.12156862745098039, 0.4666666666666667, 0.7058823529411765}
\definecolor{c2}{rgb}{1.0, 0.4980392156862745, 0.054901960784313725}
\definecolor{c3}{rgb}{0.17254901960784313, 0.6274509803921569, 0.17254901960784313}
\definecolor{c4}{rgb}{0.8392156862745098, 0.15294117647058825, 0.1568627450980392}
\definecolor{c5}{rgb}{0.5803921568627451, 0.403921568627451, 0.7411764705882353}
\definecolor{c6}{rgb}{0.5490196078431373, 0.33725490196078434, 0.29411764705882354}

\newcommand{\fcirc}[1]{\textcolor{#1}{$\bullet $}\!\!\!$\circ$}

\usepackage{mdwlist}

\usepackage{setspace}

\newcommand*{\xlinethick}[1][1.2em]{\rule[0.4ex]{#1}{1.5pt}}
\newcommand*{\xlinevthick}[1][1.2em]{\rule[0.4ex]{#1}{2.5pt}}

\newcommand*{\xlineshort}[1][1.2em]{\rule[0.4ex]{3.5pt}{0.5pt}}

\newcommand*{\xdash}[1][1.2em]{\rule[0.4ex]{2.5pt}{0.5pt} \rule[0.4ex]{2.5pt}{0.5pt}}
\newcommand*{\xdashthick}[1][1.2em]{\rule[0.4ex]{3.5pt}{1.5pt} \rule[0.4ex]{3.5pt}{1.5pt}}
\newcommand*{\xdashvthick}[1][1.2em]{\rule[0.4ex]{4.5pt}{2.5pt} \rule[0.4ex]{4.5pt}{2.5pt}}

\newcommand{\eq}[1]{ Eq.\ (\ref{#1})}

\DeclareMathOperator*{\Ptype}{P}
\DeclareMathOperator*{\Pressuretype}{\textbf{P}}

\DeclareMathOperator*{\StressIK1}{\boldsymbol{\Pressuretype\limits^{\scriptscriptstyle{IK1}}}}

\DeclareMathOperator*{\StressMOP}{\boldsymbol{\Pressuretype\limits^{\scriptscriptstyle{MOP}}}}
\DeclareMathOperator*{\PressureVA}{\boldsymbol{\Pressuretype\limits^{\scriptscriptstyle{V\!A}}}}

\DeclareMathOperator*{\StressI}{{\Ptype \limits^{\scriptscriptstyle{IK}}}}
\DeclareMathOperator*{\StressHara}{{\Ptype \limits^{\scriptscriptstyle{H}}}}
\DeclareMathOperator*{\StressSF}{{ \Ptype \limits^{\scriptscriptstyle{SF}}}}

\newcommand{\insurfi}[1]{{\xi_{i}\!}^{\textit{#1}}}

\newcommand{\insurfl}[1]{\xi_{\lambda}^{\textit{#1}}}

\newcommand{\eg}{{e.g.\ }}

\begin{document}

\title{The Importance of Reference Frame for Pressure at the Liquid-Vapour Interface}
\author{
\name{Edward R. Smith}
\thanks{edward.smith@brunel.ac.uk}
\affil{Department of Mechanical and Aerospace Engineering, Brunel University London}
}

\maketitle

\begin{abstract}

The local pressure tensor is non-unique, a fact which has generated confusion and debate in the seventy years since the seminal work by Irving Kirkwood. 
This non-uniqueness is normally attributed to the interaction path between molecules, especially in the interfacial-science community. 
In this work we reframe this discussion of non-uniqueness in terms of the location, or reference frame, used to measure the pressure.
 By using a general mathematical description of the liquid-vapour interface, we obtain a reference frame that moves with the interface through time, providing a new insight into the pressure. 
We compare this instantaneous moving reference frame with the fixed Eulerian one. 
Through this process, we show the requirement that normal pressure balance at the moving surface is satisfied by surface fluxes, however an additional corrective term based on surface curvature is required for the average pressure in a volume.
We make the case that a focus on the path of integration is the cause of much of the confusion in the literature.
Using an explicit reference frame with a more general derivation of pressure clarifies some of the issues of uniqueness in the pressure tensor and provides a pressure tensor which is defined at any instant in time and valid away from thermodynamic equilibrium.

\end{abstract}



\section{Introduction}

Since the pioneering work of \citet{Irving_Kirkwood} we have had a firm theoretical foundation for the pressure tensor in statistical mechanics. 
In practice, however, there remains an ongoing debate about what form of pressure tensor is the right one to use in a molecular dynamics (MD) simulation. 
\citet{Irving_Kirkwood} present the mathematics of a point in space, a direct consequence of using continuum quantities in a discrete system.
Mathematically this gives us a Dirac delta function, which is an infinitesimal point and so must be adapted for use in a molecular dynamics simulation, expanding it over a finite area in order to collect measurements.
Different communities of researchers have done this in different ways.
The simplest approach is to set the delta functions to one and use the tensor version of the Virial \citep{Parker} at a local point in space to get the pressure.
This is the IK1 approximation, so called because it is a first order term in the full Irving Kirkwood expressions \citep{Evans_Morris}.
In the solid mechanics community, the delta function is sometimes replaced by defining a non-infinite function with a finite width and compact support as in \citep{Hardy, Murdoch,Murdoch_2010, Admal_Tadmor}.
These solid mechanics stress tensors, where stress is simply the negative of the pressure tensor, are largely in the form of averages inside volumes, so have become known as the volume average (VA) expressions \citep{Cormier_et_al}.
For the non-equilibrium molecular dynamics (NEMD) community, an approach treating the functions in Fourier space \citep{Lutsko, todd_daivis_2017, Evans_Morris} has been followed to yield mathematically tractable forms.
The most well-known form is the method of planes (MOP) \citep{Todd_et_al_95} and its extensions \citep{todd_daivis_2017, Han_Lee}.
Both the MOP and VA forms of pressure can be shown to be equivalent in the limit of small volumes \citep{Heyes_et_al}, both demonstrating a pressure gradient of zero normal to a solid-liquid interface, a requirement for mechanical equilibrium. 
This is worth noting as the IK1 form of pressure fails to satisfy this requirement for mechanical equilibrium \citep{Varnik_et_al(2000)}.
In addition to the VA, IK1 and MOP pressures from the solid mechanics and NEMD communities, researchers interested in the interface between two fluids, in particular between a liquid and a vapour, tend to work with pressure obtained relative to that interface.
There are two common forms obtained by integrating relative to the interface area itself, called the Irving Kirkwood \citep{Walton_et_al83} and \citet{Harasima1958aa} contours.
The difference between the Harasima and Irving Kirkwood forms is attributed to the different interaction paths, or interaction contours, assumed between the molecules \citep{Rowlinson2002aa}.
This is generalised in the work of \citet{Schofield_Henderson} who show that the path of interaction between two particles, which defines their interaction force, can occurs in an infinite number of ways.
This is explored for various contours in recent work \citep{Shi_et_al21}.
Far from a philosophical point, this non-uniqueness has profound implications for measurements in MD systems. 
Published results were recently called into question due to the non-uniqueness of pressure \citep{C9CP02890K, C9CP04289J}.
In addition, use of Harasima and Irving Kirkwood surface forms led to the conclusion that Marangoni flow cannot be measured at the molecular scale \citep{Liu_et_al17}. 
\citet{Malijevsk__2012} concluded the mechanical route to pressure is unreliable and should be avoided in systems with thermal fluctuations. 
In an attempt to  address the confusion in the literature, \citet{Varnik_et_al(2000)} compared different forms of pressure, considering the Harasima and Irving Kirkwood contour along with the method of planes on a flat surface.
In this work we aim to extend this comparison to an arbitrary liquid-vapour interface, working with the VA pressure and a generalisation of the MOP pressure we call the surface flux (SF) form, we then discuss the relationship between these forms and the Harasima and Irving Kirkwood contours.

This consideration of pressure on a general interface shifts the focus away from the interaction contour between two molecules.
A focus on the contour of interaction ignores the range of other factors in defining the pressure tensor.
For example, the split of the kinetic term into streaming and fluctuating part is not free from ambiguity \citep{Zhou}, a point which takes on new importance for a reference frame moving with the interface.
In addition, the measured pressure is only correct to a gauge, meaning any divergence-free term could be added and the equations of motion would still hold \citep{Todd_et_al_95}.
Most importantly for interfaces, both the interaction contour and the location of measurement are required to determining the pressure, a point highlighted in the original work of \citet{Irving_Kirkwood} who note the force acting and the location it acts across, $dS$, are quite arbitrary.
In comparing Harasima and Irving Kirkwood forms, \citet{Varnik_et_al(2000)} note the interactions will be counted upon crossing an area, showing the contour path definitions are intertwined with the location of measurement.
A contour could therefore be chosen which avoids being counted because it misses the area used to measure pressure.
Here the SF formulation has a clear advantage, as it measures pressure on six connected surfaces enclosing a volume, for a molecule in the volume it is impossible for any choice of contour to not cross at least one surface.
The location of SF pressure measurment is therefore not a plane but a volume defined by the set of connected patches which encloses it.
The measurement location is typically defined to be a located at the centre of the liquid-vapour interface \citep{Walton_et_al83}, the surface of a sphere \citep{Walton_et_al85, Hafskjold_Ikeshoji02}, cylinder \citep{Shi_et_al20}, and so on.
A number of recent publications have shown we can track the liquid-vapour interface instantaneously down to the intermolecular spacing, \citep{Chacon2003aa, Partay2008ab, Willard2010aa, Jorge2010ab, Sega2013aa}, moving with the molecules as they evolve in time.
Instead of considering the average  of a plane (or sphere, cylinder, etc), in this work we use a pressure measurement which follows the instantaneous surface described by an arbitrary function $\xi = \xi(x,y,t)$.
This is mathematically tedious, but can be done for both VA \citep{Braga_et_al18} and  SF pressure  \citep{Smith_Braga20}.
Using a measurement relative to an arbitrary surface gives the most general possible form of interface pressure, where $\xi$ could then be chosen to be a flat surface, spherical segment, etc.
As a result, these are valid for any surface away from equilibrium at every timestep.
We provide a comparison of the resulting profiles for these SF pressures to the expression obtained inside a volume (IK1, VA) using both a flat plane and one moving with the instantaneous surface.
Results for both the solid-liquid interface and liquid-vapour interface are compared and it is shown that Harasima and Irving Kirkwood contours are a special case of the more general VA and SF pressures.
Through this comparison, the importance of explicit consideration of the reference frame is highlighted.
The use of a reference frame fluctuating with the instantaneous interface, changes the pressure definition from thermodynamic to purely mechanical by accounting for every single force and flux. 
With every crossing counted, measured SF pressures on all surfaces of a volume can therefore be shown to exactly balance the momentum change in the volume, to machine precision \citep{Smith_Braga20}.
This addresses at least one concern of thermal fluctuations invalidating the pressure measurement \citep{Malijevsk__2012}.
The long-range contributions \citep{Sega_et_al16} as well as three body \citep{todd_daivis_2017} and greater interactions are not considered in this work, although the exactly balance on a control volume might prove useful in deriving these cases.



This manuscript is structured as follows, we start with an overview of the various mathematical forms of the pressure tensor in section \ref{sec:theory}, with a novel correction term for the VA pressure derived at the end. Next, the molecular setup is discussed, for the two cases simulated in this work, a solid-liquid interface for reference and the liquid-vapour interface in section  \ref{sec:methods}. The results and discussion from these cases are presented next in section \ref{sec:results}, with particular focus on the results taken in a reference frame moving with the interface, followed by the conclusions in section \ref{sec:conclusions}.

\section{Theory}
\label{sec:theory}

We briefly present the theoretical developments that lead to the different forms of pressure tensor.

\subsection{Irving Kirkwood Pressure Tensor}

The derivation of the pressure tensor is given in the work of \citet{Irving_Kirkwood}, taking the time evolution of momentum and comparing  forms with the expected continuum equations.
The resulting pressure tensor at point $\boldsymbol{r}$ in the fluid is of the form,
\begin{align}
\boldsymbol{P}(\boldsymbol{r},t) = \displaystyle\sum_{i=1}^{N} \bigg\langle  m_i\left(\dot{\boldsymbol{r}}_i-\boldsymbol{u} \right) \left( \dot{\boldsymbol{r}}_i - \boldsymbol{u} \right) \delta(\boldsymbol{r}_i-\boldsymbol{r}) 
 +  \frac{1}{2}   \displaystyle\sum_{j\neq i}^{N}  \boldsymbol{f}_{ij}  \boldsymbol{r}_{ij}  O_{ij}\delta(\boldsymbol{r}_i-\boldsymbol{r}) ; \textit{f}  \bigg\rangle.
\label{IKstress}
\end{align}
where angular brackets $\langle a ; f\rangle$ denote the phase space average of $a$, velocity of molecule $i$ is given by $\dot{\boldsymbol{r}}_i$ and the streaming velocity is $\boldsymbol{u}$.
The distance between the positions of molecules $i$ and $j$ is defined by $\boldsymbol{r}_{ij} = \boldsymbol{r}_i-\boldsymbol{r}_{j}$ and $\boldsymbol{f}_{ij} = -\partial \phi_{ij} / \partial r_{ij}$ denotes the inter-molecular force (the force exerted by $i$ on $j$). 
The first term is the kinetic part of the pressure tensor, which includes the contributions due to molecular velocity. The second term is the configurational pressure, due to the interactions between molecules. 
The $O_{ij}$ term is known as the IK operator \citep{Todd_et_al_95}, obtained from the difference between the Dirac delta functions for molecule $i$ and $j$,
\begin{align}
\delta(\boldsymbol{r}_i-\textbf{r}) - \delta(\boldsymbol{r}_j-\boldsymbol{r})=  - \boldsymbol{r}_{ij} \cdot \frac{\partial}{\partial \boldsymbol{r}} O_{ij} \delta(\boldsymbol{r}_i-\boldsymbol{r}),
\label{diff_dd}
\end{align}
where $O_{ij}$ denotes a Taylor expansion operator,
\begin{align}
O_{ij} = \left( 1 - \frac{1}{2}\boldsymbol{r}_{ij} \cdot \frac{\partial}{\partial \boldsymbol{r}_i} + \ldots + \frac{1}{n!}(-\boldsymbol{r}_{ij} \cdot \frac{\partial}{\partial \boldsymbol{r}_i} )^{n-1} + \ldots \right)
\label{IKOij_1}
\end{align}
In what follows, we drop the phase space average so we can obtain expressions that are valid instantaneously in a molecular dynamics simulation. Introducing the definition $\boldsymbol{p}_{i}/m_i =  \dot{\boldsymbol{r}}_i-\boldsymbol{u} $, \eq{IKstress} therefore becomes,
\begin{align}
\boldsymbol{P}(\boldsymbol{r},t) = \displaystyle\sum_{i=1}^{N}  \frac{\boldsymbol{p}_i \boldsymbol{p}_{i}}{m_i} \delta(\boldsymbol{r}_i-\boldsymbol{r}) 
 +  \frac{1}{2}   \displaystyle\sum_{i,j}^{N}  \boldsymbol{f}_{ij}  \boldsymbol{r}_{ij}  O_{ij}\delta(\boldsymbol{r}_i-\boldsymbol{r}) .
\label{IKstress_instant}
\end{align}
Where the sum notation $\sum_{i,j}^{N} = \sum_{i=1}^{N} \sum_{j \ne j}^{N}$ has been introduced for conciseness. 

\subsection{Irving Kirkwood One (IK1) Tensor}

First we consider \eq{IKstress_instant} with only the first term in the IK operator of \eq{IKOij_1}, know as the IK1 approximation. The result is the tensorial form of the virial pressure \citep{Parker} but applied locally,
\begin{align}
\StressIK1(\boldsymbol{r},t) = \displaystyle\sum_{i=1}^{N} \frac{\boldsymbol{p}_i \boldsymbol{p}_{i}}{m_i}  \delta \left( \boldsymbol{r} - \boldsymbol{r}_i\right)   + \frac{1}{2} \displaystyle\sum_{i,j}^{N} \boldsymbol{f}_{ij} \boldsymbol{r}_{ij} \delta \left( \boldsymbol{r} - \boldsymbol{r}_i\right).
\label{IK1}
\end{align}
The IK1 form of pressure is the most widely used in the molecular dynamics literature due to its simplicity, and at the time of writing is the default in LAMMPS, calculated from a binning of the per-atom stress \citep{PLIMPTON19951}. 
Using the virial pressure locally in an inhomogeneous system is incorrect, as interactions with the surrounding fluid are not included \citep{Tsai}. 
This can also be interpreted as a consequence of neglecting terms of higher order in \eq{IKOij_1} so effects of local inhomogeneity in the fluid are lost. 
Away from equilibrium a localised description is required and the full \eq{IKOij_1} expression must be retained.

%
%
%
%
%
%
%
%
\subsection{The Contour Forms - Irving Kirkwood and Harasima}
In order to retain the higher order terms of \eq{IKOij_1}, the delta functions in \eq{diff_dd} can be reformulated as a contour integral between the two molecules.
In order to do this, we rewrite the IK operator using the fundamental theorem of contour integration,
\begin{align}
 \delta({\bf{r}} - {\bf{r}}_i) -  \delta({\bf{r}} - {\bf{r}}_j) 
&= \oint \frac{\partial}{\partial \boldsymbol{\ell}}  \cdot \delta \left({\bf{r}} - \boldsymbol{r}_i - \boldsymbol{\ell} \right)  d\boldsymbol{\ell} \nonumber \\
&=  \oint  \frac{\partial \bf{r}_i + \boldsymbol{\ell} }{\partial \boldsymbol{\ell}}  \cdot \frac{\partial}{\partial \left[{\bf{r}}_i + \boldsymbol{\ell} \right]}\delta \left({\bf{r}} - \boldsymbol{r}_i - \boldsymbol{\ell} \right)  d\boldsymbol{\ell}
 \nonumber \\
&= \frac{\partial}{\partial {\bf{r}}} \cdot  \oint  \delta \left({\bf{r}} - \boldsymbol{r}_i - \boldsymbol{\ell} \right)  d\boldsymbol{\ell},
\label{General_contour}
\end{align}
where the integral along $\boldsymbol{\ell}$  represents an infinite number of paths of integration between the atoms $i$ and $j$ \citep{Schofield_Henderson}.
In the classical paradigm of molecular dynamics, especially for the pairwise potentials considered in this work, Occam's Razor dictates that the interaction is a straight line between molecules.
This is also in line with Newton's assumption of impressed force between two points.
This assumptions results in $\boldsymbol{\ell} = \lambda \boldsymbol{ r_{ij}}$ with $0 < \lambda < 1$,
\begin{align}
 \frac{\partial}{\partial {\bf{r}}} \cdot  \oint  \delta \left({\bf{r}} - \boldsymbol{r}_i - \boldsymbol{\ell} \right)  d\boldsymbol{\ell} = \frac{\partial}{\partial {\bf{r}}} \cdot  \boldsymbol{r}_{ij} \int_{0}^{1}  \delta \left({\bf{r}} - \boldsymbol{r}_i - \lambda \boldsymbol{ r_{ij}} \right)   d\lambda ,
 \label{contoureq}
\end{align}
where $d\boldsymbol{\ell} = \boldsymbol{ r_{ij}} d\lambda $.
so the configurational pressure is,
\begin{align}
\boldsymbol{P}^C(\boldsymbol{r},t) =  \frac{1}{2}   \displaystyle\sum_{i,j}^{N}  \boldsymbol{f}_{ij}  \boldsymbol{r}_{ij} \int_{0}^{1}  \delta \left({\bf{r}} - \boldsymbol{r}_i - \lambda \boldsymbol{ r_{ij}} \right)   d\lambda  .
\label{Config_pressure}
\end{align}
Now as $\delta \left({\bf{r}} - \boldsymbol{r}_i - \lambda \boldsymbol{ r_{ij}} \right) = \delta \left(x- x_i - \lambda x_{ij} \right) \delta \left(y- y_i - \lambda y_{ij} \right) \delta \left(z- z_i - \lambda z_{ij} \right)$ , the Irving Kirkwood contour can be seen to be the special case when delta functions in the $x$ and $y$ directions are set to one and we get the resulting form of pressure over a flat interface in $z$ by integrating along the line of interaction,
\begin{align}
\int_{0}^{1}  \delta \left(z - z_i - \lambda z_{ij} \right)   d\lambda  = \frac{1}{|z_{ij}|} H\left( \frac{z - z_i}{z_{ij}} \right) H\left( \frac{z_j - z}{z_{ij}} \right),
\end{align}
with $H$ the Heaviside function.
This choice of a straight line contour in $z$, crossing a fixed area, appears in the appendix of the original paper by \citet{Irving_Kirkwood} to derive,
\begin{align}
\boldsymbol{\StressI{^C}}(\boldsymbol{r},t) =  \frac{1}{2}   \displaystyle\sum_{i,j}^{N}  \boldsymbol{f}_{ij}   \frac{\boldsymbol{r}_{ij}}{|z_{ij}|} H\left( \frac{z - z_i}{z_{ij}} \right) H\left( \frac{z_j - z}{z_{ij}} \right),
\label{Config_IK_pressure}
\end{align}
with components,
\begin{align}
{\StressI{^C_N}} & = \StressI{_{zz}^C} =  \frac{1}{2}   \displaystyle\sum_{i,j}^{N}  {f}_{zij}   \frac{{z}_{ij}}{|z_{ij}|} H\left( \frac{z - z_i}{z_{ij}} \right) H\left( \frac{z_j - z}{z_{ij}} \right), 
\label{IK_Normal} \\
{\StressI{^C_T}} &  = \frac{1}{2} \left[\StressI{_{xx}^C} + \StressI{_{yy}^C} \right]=  \frac{1}{4}   \displaystyle\sum_{i,j}^{N} \left[    \frac{ {f}_{xij}{x}_{ij} +  {f}_{yij}{y}_{ij}}{|z_{ij}|} \right] H\left( \frac{z - z_i}{z_{ij}} \right) H\left( \frac{z_j - z}{z_{ij}} \right), 
\label{IK_Tangential}
\end{align}
This form of the pressure is known in the literature as the Irving Kirkwood contour \citep{Walton_et_al83}.
It has been shown that the normal pressure \eq{IK_Normal} is equivalent to the normal part of the MOP expression \citep{Varnik_et_al(2000)}, however, the tangential parts of \eq{IK_Tangential} appears to be a hybrid of the VA and MOP forms.
The pressure expression $ {f}_{xij}{x}_{ij} $ of \eq{IK_Tangential} is the same as the tangential VA component but counted like the normal MOP contribution, when a crossing occurs on the $z$ plane. 

The other common formulation of pressure in the interface literature is based on the work of \citet{Harasima1958aa}, which adjusts the interaction contour between the molecules based on the location of the interface of interest.
The contour splits the path into a tangential and a normal component to a surface, so for a flat interface, the resulting form of stress is, 
\begin{align}
\StressHara{_N} & = \StressI{_N}  \\
\StressHara{_T} & =   \frac{1}{4}   \displaystyle\sum_{i,j}^{N} \left[    \frac{ {f}_{xij}{x}_{ij} +  {f}_{yij}{y}_{ij}}{|z_{ij}|} \right]\delta( z - z_i).
\end{align}
For more complicated surfaces, with position in the $z$ direction described by a function $\xi = \xi(x,y,t)$, the normal $\boldsymbol{n} = \boldsymbol{\nabla} \xi / |\boldsymbol{\nabla} \xi|$ and tangents $\boldsymbol{t_1}$ and $\boldsymbol{t_2}$ would be used to give an interaction path moving along the tangential and then normal direction.
The general contour of \eq{General_contour} can be considered to be $\boldsymbol{\ell} = \boldsymbol{\ell}_t + \boldsymbol{\ell}_n$ with $\boldsymbol{\ell}_t$ and $\boldsymbol{\ell}_n$ defined as tangent and normal to the interface function $\xi$.
This choice of contour has a side effect. 
The inter-molecular interaction between the molecules is based on the current shape of the liquid-vapour interface and the reference frame we fit to it.
It is argued here that it is more natural to consider the inter-molecular interaction to be the same regardless of system and instead explicitly consider the reference frame changing based on the interface.
We will then obtain expressions which are similar to the Harasima contour but derived in terms of the reference frame.
To do this, we integrate over a volume as in \citet{Lutsko} and \citet{Cormier_et_al}, which leads to the so-called Volume Average (VA) pressure formulation.

\subsection{Volume Average (VA) Pressure}

The volume average (VA) pressure is obtained from a volume integral of \eq{IKstress_instant}, assuming linear inter-molecular interactions using Eqs. (\ref{General_contour}) and (\ref{contoureq}).
The location, size and shape of the volume is arbitrary and can be chosen wherever we want to measure the pressure.
The kinetic pressure includes the kinetic contributions of molecules inside the volume, while the configurational term is based on the fraction of an inter-molecular interaction that passes through the given volume in space. 
This has the advantage that it is equivalent to the virial when all volumes in space are added up, as all fractions are included,
\begin{align}
\PressureVA = \frac{1}{\Delta V} \left[ \displaystyle\sum_{i = 1 }^{N}  \frac{\boldsymbol{p}_i \boldsymbol{p}_{i}}{m_i}  \vartheta_i + \frac{1}{2}\displaystyle\sum_{i,j}^{N}    \boldsymbol{f}_{ij} \boldsymbol{r}_{ij} \int_{0}^{1} \vartheta_{\lambda} d \lambda  \right],
\label{VA_stress}
\end{align}
where $\Delta V$ is the local volume; the $\vartheta_i$ function is only non-zero for a molecule inside the volume and the $\vartheta_{\lambda}$ is non zero when a fraction of the interaction is inside the averaging volume.
The $\vartheta$ function is simply the integral of the Dirac delta function over a volume \citep{Smith_et_al12, Heyes_et_al14, Braga_et_al18}, $\vartheta_{\alpha} = \Lambda_{x \alpha} \Lambda_{y \alpha} \Lambda_{z \alpha}$ where $\Lambda_{x \alpha}  = H(x^+ -x_{\alpha}) - H(x^- - x_{\alpha})$ with $H$ the Heaviside function, $x^+$ the top of a volume in $x$, $x^-$ the bottom and $x_{\alpha}$ the position of either a molecules when ${\alpha}=i$ or a point on the line of interaction when ${\alpha}=\lambda$ so $x_{\lambda} = x_i + \lambda x_{ij}$.
The difference between two Heaviside functions is known as a boxcar function, shown in figure \ref{schematic_surface}, a function which is one between two points and zero outside. 

An advantage of the VA approach is we can choose any volume, including one which is moving with the interface $\xi$ in the $z$ direction, so this $z$ boxcar function becomes $\Lambda_{z \alpha}  = H(z^+ + \xi(x_{\alpha}, y_{\alpha}, t) - z_{\alpha}) - H(z^- + \xi(x_{\alpha}, y_{\alpha}, t) - z_{\alpha})$. 
Mathematically the value of $z_{\alpha}$, with molecule ${\alpha}=i$ or line ${\alpha}=\lambda$, can be interpreted as shifted based on the surface $\xi$ at the same $x_{\alpha}$ and $y_{\alpha}$ location
, where this form follows directly from the mathematical derivation \citep{Braga_et_al18}.
The shape of a typical volume with the $z$ surface following the interface can be seen in figure \ref{schematic_surface}.
As a result, it has some similarities to the Harasima approach, expressing terms parallel to the current interface. 
We can simplify the integral along $\lambda$ in the Harasima case by noting that if molecule $i$ is in the volume, the integration of a contour over all components tangential to the surface must also be inside.
Assuming the integral of the $\Lambda_{\lambda x}$ and $\Lambda_{\lambda y}$ functions give the line segment in the volume in $x$, which is $\Delta x$, and in $y$, which is $\Delta y$, and taking $P_T^C$ as the sum of the $xx$ and $yy$ components, 
\begin{align}
P_T^C =  \frac{1}{4}\displaystyle\sum_{i,j}^{N}  \left[ \frac{  {f}_{x ij} {x}_{ij}  +  {f}_{y ij} {y}_{ij}}{ \Delta z}\right] \Lambda_{iz} ,
\label{VA_Harasima_stress}
\end{align}
This is similar to the tangential pressure part of the Harasima contour, but derived assuming the reference frame is moving with the surface.

Finally, it is worth mentioning that nothing in the integration process over a volume depends on the contour, so we could write the general form of contour integral,
\begin{align}
\PressureVA = \frac{1}{\Delta V} \left[ \displaystyle\sum_{i = 1 }^{N}  \frac{\boldsymbol{p}_i \boldsymbol{p}_{i}}{m_i}  \vartheta_i + \frac{1}{2}\displaystyle\sum_{i,j}^{N}    \boldsymbol{f}_{ij} \boldsymbol{r}_{ij} \oint \vartheta_{\boldsymbol{\ell}} d \boldsymbol{\ell} \right],
\label{VA_stress_general}
\end{align}
where the length of a contour in a volume is obtained from the integral over $ \vartheta_{\boldsymbol{\ell}} = \Lambda_{x \ell} \Lambda_{y \ell} \Lambda_{z \ell}$ function with $\Lambda_{x \ell} = H(x^+ -x_i - \ell_x) - H(x^- - x_i - \ell_x)$  where $\ell_x$ is the $x$ component of the contour, similarly for $y$ with the moving $z$ surface, $\Lambda_{z \ell} = H(z^+ + \xi(\ell_x, \ell_y,t) -z_i - \ell_z) - H(z^- + \xi(\ell_x, \ell_y,t) -z_i - \ell_z)$.
In practice, this would be obtained in a piecewise manner in a computer code, stepping along a contour and evaluating $\vartheta_{\boldsymbol{\ell}}$ for each segment.
Different contours would give different results, for example a few contours are given in Figure \ref{schematic_surface}, which each show different lengths inside the volume.
The next section expresses pressure as surface crossings, which avoids this ambiguity, counting contributions entirely based on which surface is crossed.

\subsection{Method of Planes (MOP) and Surface Flux (SF) Pressure}
The surface pressure approach avoids the expansion in delta functions of \eq{IKOij_1} by considering the interaction over a plane; introduced empirically by \citet{Tsai} and derived formally by a Fourier transform of the Irving Kirkwood expression, \eq{IKstress} in \citet{Todd_et_al_95}. 
This MOP form has the ability to deal with systems arbitrarily far from equilibrium \citep{Todd_et_al_95}. 
It also gives three components of the stress tensor acting across a plane, when the plane normal is aligned along the $x$ axis this is,
 \begin{align}
\StressMOP{\!\!_x}   = \frac{1}{\Delta A_x} \left[\displaystyle\sum_{i = 1 }^{N} \frac{\boldsymbol{p}_i p_{ix}  }{m_i} \delta(x_i-x)  + \frac{1}{4} \displaystyle\sum_{i,j}^{N} \boldsymbol{f}_{ij} \left[sgn(x - x_j) - sgn(x - x_i)\right] \right] .
 \end{align}
where $sgn(x)$ is the signum function. 
This two signum functions have the interpretation of including inter-molecular crossings over the surface of the plane.
The normal component $ P_{xx}$ uses the normal components ${p}_{ix}$ and $f_{xij}$, and is equivalent to the IK form of pressure in the normal direction given in \eq{IK_Normal} (as $sgn(z_{ij}) = z_{ij}/|z_{ij}|$ is cancelled by the $z_{ij}$ denominator in \eq{IK_Normal}, see \citep{Smith_et_al12}). 
The other two components of momentum and force provide the two other pressure components, $ P_{xy} \text{ and } P_{xz} $, and returns a single value for a plane covering the whole domain. 
\citet{Han_Lee} used three mutually perpendicular planes converging at a point to obtain all nine components of stress, with planes limited to a local region of interest. 
This allows us to write the tangential components of pressure $ P_{yy}, P_{yx} \text{ and } P_{yz} $ as,
 \begin{align}
\StressMOP{\!\!_y}    = & \frac{1}{\Delta A_x} \Bigg[\displaystyle\sum_{i = 1 }^{N} \frac{\boldsymbol{p}_i p_{ix}  }{m_i} \delta(y_i-y) \Lambda_{x i} \Lambda_{z i}   
\nonumber \\ 
& + \frac{1}{4} \displaystyle\sum_{i,j}^{N} \boldsymbol{f}_{ij} \left[sgn(y - y_j) - sgn(y - y_i)\right] \Lambda_{x }(\lambda_k) \Lambda_{z} (\lambda_k) \Bigg] .
\label{Han_Lee_MOP}
 \end{align}
where $\Lambda_{x i}$ and $\Lambda_{x } (\lambda_k)$ are boxcar functions which localise the plane to a patch in space. 
These check if the molecular $i$ and point of crossing $(\lambda_k)$ are between the limits of the surface in $x$ (similar for $y$) with $\Lambda_{x } (\lambda_k) = H(x^+ - x_i + \lambda_k x_{ij}) - H(x^- - x_i + \lambda_k x_{ij})$.
It is instructive to compare this pressure, where $P_{yy}$ is the $y$ component of force on the $y$ surface, with the Irving Kirkwood contour of \eq{IK_Tangential} which measures the $y$ force times $y_{ij}$ but on the $z$ surface.
This difference will be small for thin volumes but could become significant as $\Delta z$ increases.

\begin{figure}
  \centering
    \includegraphics[width=1.0\textwidth]{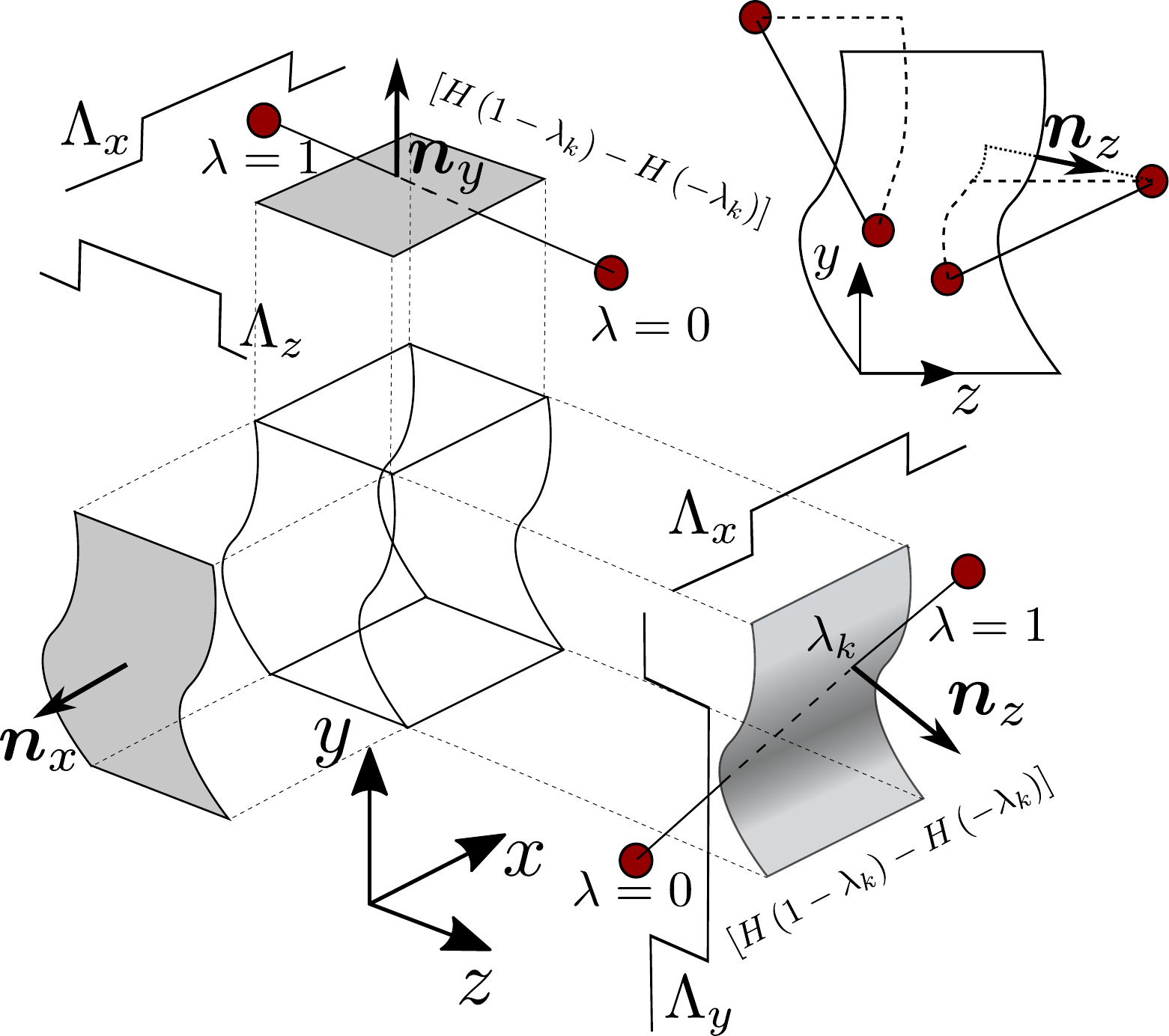}
      \caption{A schematic showing a single control volume and the action of the different mathematical terms in the SF expression. The location of a surface crossing, $\lambda_k$, is checked to be between the two molecules using the difference $H(1-\lambda_k) - H(-\lambda_k)$ while the various $\Lambda$ expression act to limit the crossing to within a rectangular region of the surface. The normal at the point of crossing is used in the calculation of the surface term. The top right shows different example contours between molecules including the Irving Kirkwood as a solid line and two interpretations of the Harasima as different dotted lines assumed to move tangentially to the surface until it gets to a line either normal (which is consistent with the definition but can have multiple solutions for a arbitrary surface) or along the $z$ axis}
      \label{schematic_surface}
\end{figure}

The local MOP pressure proposed by \citet{Han_Lee} occurs naturally from the derivation in the original paper by \citet{Irving_Kirkwood}, by working in integrated, or control volume, form \citep{Smith_et_al12}.
This proceeds by integrating the \citet{Irving_Kirkwood} momentum over a volume and evaluating the time evolution of this volume \citep{Smith_et_al12}.
The resulting equations have a number of advantages, the pressure on the six volume surfaces can be used to get all nine components as in the \citet{Han_Lee} form.
The derivation provides a natural link between VA and MOP pressures \citep{Smith_et_al12} while the use of a closed volume means any choice of contour \textbf{\textit{must}} be included on one of the surfaces (see Figure \ref{schematic_surface}).
This directly addresses the non-uniqueness of interaction contour, as all interactions must be included provided one molecule is inside the volume and the other outside, so contour choice simply changes which surface an interaction is counted on.
This has the effect of shifting it from a shear pressure contribution to a direct contribution (or vice versa), and becomes similar to choosing a different slice through a material.
A notable corollary of this is that while $\boldsymbol{\nabla} \cdot \boldsymbol{P} = 0$ is satisfied, shear components \eg $\partial P_{xy} / \partial x$ may be non-zero for certain choices of contour.
We can therefore verify direct components are constant and shear components are zero (required in a Newtonian liquid by definition), as a test that the choice of contour is meaningful.
The other important advantage of the control volume approach is by defining the surfaces enclosing a volume, the sum over all surfaces is exactly equal to momentum change inside the volume. 
The pressure equilibrium condition is generalised to a form valid every timestep, so becomes $\boldsymbol{\nabla} \cdot \boldsymbol{P} = d\rho \boldsymbol{u}/dt  + \boldsymbol{\nabla} \cdot \rho \boldsymbol{u}\boldsymbol{u}$ in control volume form.
This allows us to derive a general surface pressure expression we are sure includes all terms for surface movement and curvature, by checking the sum of forces and momentum on a volume balance exactly.
It is this property that has been used to validate the result for pressure presented in this work.

In order to generalise the MOP form to a surface, we introduce the notation $dS_{xi}^+ = \delta(x_i-x^+) \Lambda_{i y} \Lambda_{i z}$ and $dS_{x\lambda}^+ = \delta(x_\lambda - x^+), \Lambda_{\lambda y} \Lambda_{\lambda z}$ for the top surface (denoted by the $+$ superscript) in the $x$ direction,
  \begin{align}
  \rho \boldsymbol{u} u_x + \boldsymbol{\StressSF{_x^+}} = \frac{1}{\Delta S_x}  \;\;\; \displaystyle\sum_{i=1}^N  m_i \boldsymbol{\dot{r}}_i \dot{x}_i dS_{xi}^+ + \frac{1}{2 \Delta S_x}\sum_{i,j}^{N} \boldsymbol{f}_{ij}  
\int_0^1   {x}_{ij}  dS_{x\lambda}^+  d\lambda,
\label{x_stress}
 \end{align}
 and it can be shown that,
  \begin{align}
   \int_0^1   {x}_{ij}  dS_{x\lambda}^+  d\lambda = \left[sgn(x^+ - x_j) - sgn(x^+ - x_i)\right]  \Lambda_{y} (\lambda_k) \Lambda_{z} (\lambda_k)
  \end{align}
for a flat interface.
The use of peculiar momentum $\boldsymbol{p}_i$ has been dropped and the convective terms moved to left-hand side to avoid making the assumption that streaming velocity is constant.
Equation (\ref{x_stress}) can therefore be seen to be the localised MOP shown in \eq{Han_Lee_MOP}, which we call the Surface Flux (SF) pressure, because it can be generalised to get fluxes on any surface, including ones which are not flat.
The $\Lambda_{\alpha z}$ term is a function of the moving surface  $\xi$, so the size of the rectangular part of the plane it selects is constant, but moves as the surface changes.
As a result, it is always the same distance from the surface and the same width (see Fig \ref{schematic_surface}).

For interactions crossing the arbitrary surface itself, $\xi$, the SF pressure is expressed as follows,
 \begin{align}
  \rho \boldsymbol{u} u_z + \StressSF{_z^+}  
& = \frac{1}{\Delta S_z}  \;\;\; \displaystyle\sum_{i=1}^N  m_i \boldsymbol{\dot{r}}_i \;\;\; \Big [\overbrace{\dot{x}_i \frac{\partial \insurfi{+}}{\partial x_i} \; + \; \dot{y}_i \frac{\partial \insurfi{+}}{\partial y_i}}^{\textit{Kinetic Curvature}} \;  + \; \dot{z}_{i}
+  \overbrace{  \frac{\partial \insurfi{+}}{\partial t} \vphantom{\frac{\partial \insurfi{+}}{\partial y_i}}}^{\textit{\!\!\!\!\!\!\!\!\!\!\!\!\!\!\!Surface Evolution\!\!\!\!\!\!\!\!\!\!\!\!\!\!\!}}  \Big] dS_{zi}^+
\nonumber  \\ 
& + \frac{1}{2 \Delta S_z}\sum_{i,j}^{N} \boldsymbol{f}_{ij}  
\int_0^1 \Big[ \underbrace{ {x}_{ij} \frac{\partial \insurfl{+} }{\partial x_\lambda} + {y}_{ij} \frac{\partial \insurfl{+} }{\partial y_\lambda} }_{\textit{Configurational Curvature}} \! + \, {z}_{ij}  \Big]  dS_{z\lambda}^+  d\lambda, \;\;\;\;\;\;\;\;\;\;\;\;\;\;\;\;\;
\label{z_stress}
\end{align}
where here the $dS_{z\alpha}^+ = \delta(\xi_{\alpha}^+ - z_\alpha)  \Lambda_x  \Lambda_y$ with $\xi_{\alpha}^+ =  z^+ + \xi(x_{\alpha}, y_{\alpha}, t)$ catching the point of crossing if it is on the interface surface.
We have a number of additional terms due to local surface curvature at the location of crossing.
These include the crossing of a molecular trajectory and the crossing of an intermolecular interaction for kinetic and configurational parts respectively, as well as a surface time-evolution term.

For the case when $\xi$ is constant or zero, the expression of \eq{z_stress} will simplify to the localised form of the MOP pressure given in \eq{x_stress}, so this expression can be used to get pressure on every surface of a volume.
The different surfaces of a volume moving with an interface are shown in Figure \ref{schematic_surface}.
The expression to actually calculate the pressure on a surface in an MD simulation, here the $z^+$ surface, is, 
\begin{align}
\int_{t_1}^{t_2}  \rho \boldsymbol{u} u_z + \StressSF{_z^{k+}}  dt
 = & \frac{1}{\Delta S_z}  \displaystyle\sum_{i=1}^N  m_i \boldsymbol{\dot{r}}_i  
 \frac{\boldsymbol{r}_{12} \cdot {\boldsymbol{n}_z} }{|\boldsymbol{r}_{12} \cdot {\boldsymbol{n}_z} |} dS^+ + \frac{1}{\Delta S_z} \displaystyle\sum_{i=1}^N  m_i \boldsymbol{\dot{r}}_i  \vartheta_t 
\nonumber  \\ 
 \StressSF{_z^{c+}}= & \frac{1}{2 \Delta S_z}\sum_{i,j}^{N} \boldsymbol{f}_{ij}  
\frac{\boldsymbol{{r}}_{ij} \cdot {\boldsymbol{n}_z} }{|\boldsymbol{r}_{ij} \cdot {\boldsymbol{n}_z}|}dS^+
\label{z_equation_implementation}
 \end{align}
where $\boldsymbol{r}_{12} = \boldsymbol{r}_2 - \boldsymbol{r}_1$ is the line of time evolution of a molecule between $t$ and $t+\Delta t$, $\vartheta_t = \left[H(\xi(t) - z_i(t)) - H(\xi(t - \Delta t) - z_i(t)) \right] \Lambda_{i x} \Lambda_{i y}$ counting how many molecules have left or entered the volume due to surface movement between time $t-\Delta t$ and $t$ and the surface normal includes all of the surface curvature terms of \eq{z_stress} with $ {\boldsymbol{n}}_z = \frac{ \boldsymbol{\nabla}_\alpha  \left(   \xi - z_\alpha \right)  }{|| \boldsymbol{\nabla}_\alpha  \left(   \xi - z_\alpha \right) ||}$.
This normal is the one used in the Harasima contour for a general surface, which could be combined with tangential contour of \eq{VA_Harasima_stress}.
We have generalised the treatment of both kinetic and configurational terms in \eq{z_equation_implementation} by recognising that molecules must move from a starting point $\boldsymbol{r}_1 = \boldsymbol{r}_i(t)$ to an end point $\boldsymbol{r}_2 = \boldsymbol{r}_i(t+\Delta t)$ for surface crossings to occur. 
These movements can be written as an integral between two points in space (the molecule at different times), which makes it identical in form to the configurational term, an integral between two points (different molecules) in space.
Both kinetic and configuration terms are treated identically, the crossing of a moving molecule or the crossing of an interaction are mathematically and algorithmically equivalent.
%
In the case of a straight line and general surface $\xi$ here, this becomes a ray-tracing problem which is common in computer graphics.
Once the location of crossing, $\lambda_{k}$, is obtained, it can be inserted in the following expression,
\begin{align}
dS^+= \displaystyle\sum_{k=1}^{N_{roots}}  \left[ H \left( 1- \lambda_{k} \right) - H \left(  - \lambda_{k} \right) \right] \Lambda_x(\lambda_k)  \Lambda_y(\lambda_k),
\end{align}
with the first two Heaviside functions selecting if the root  $\lambda_k$ is between the endpoints $\boldsymbol{r}_1$ and $\boldsymbol{r}_2 $ . 
The $\Lambda$ functions then check if that crossing is in a rectangle area of the surface, see Figure \ref{schematic_surface} for a graphical illustration of this.

Multiple crossings are possible as the surface function $\xi$ is general.
Indeed, the form of crossing need make no assumptions about interaction contour or shape of surface, simply requiring a surface crossing $\lambda_k$ to be determined to evaluate $dS^+$.
It seems reasonable that this approach could also be applied to get the intersection of any contours, of the form given in \eq{General_contour}, with any surface.
The pressures tensors defined from a closed volume of bounding surfaces could then take advantage of the property already discussed which ensures any missed contributions from differing contour paths would be included as shear components on another surface.

\subsection{A Correction to the Volume Average}
\label{sec:VA_correction}
In this section we present three extra terms which must be included to get the correct form of VA pressure on a moving interface.
The VA pressure \eq{VA_stress} is commonly obtained by integrating \eq{IKstress_instant} over a volume \citep{Cormier_et_al, Braga_et_al18} .
However, we derive the SF equations from the time evolution of momentum inside a control volume,
\begin{align} 
\frac{d}{d t} \displaystyle\sum_{i=1}^N  m_i \boldsymbol{\dot{r}}_i  \vartheta_i  = \displaystyle\sum_{i=1}^N  m_i \boldsymbol{\dot{r}}_i \frac{d \vartheta_i}{d t}  + \displaystyle\sum_{i,j}^N \boldsymbol{f}_{ij} \boldsymbol{r}_{ij} \cdot 
\int_0^1    \frac{\partial \vartheta_\lambda }{\partial \boldsymbol{r}_\lambda}d \lambda
\label{mom_timevo}
\end{align}
 which results in extra terms for both the time evolution of the moving volume and the surface curvature.
 For the first term,
\begin{align} 
\displaystyle\sum_{i=1}^N  m_i \boldsymbol{\dot{r}}_i \frac{d \vartheta_i}{d t}  = \frac{\partial}{\partial \boldsymbol{r}} \cdot  \overbrace{ \displaystyle\sum_{i=1}^N  m_i \boldsymbol{\dot{r}}_i \boldsymbol{\dot{r}}_i \vartheta_i}^{\PressureVA {\!\!^k}  \Delta V } + \overbrace{ \displaystyle\sum_{i=1}^N  m_i \boldsymbol{\dot{r}}_i  \boldsymbol{\dot{r}}_i \cdot \left[ \frac{\partial \xi^+_i}{\partial \boldsymbol{r}_i} dS_{zi}^+   -  \frac{\partial \xi^-_i}{\partial \boldsymbol{r}_i} dS_{zi}^-  \right] }^{\textrm{Extra Kinetic Curvature (KC) Terms} }  
\nonumber \\
+ \underbrace{\displaystyle\sum_{i=1}^N  m_i \boldsymbol{\dot{r}}_i \left[\frac{\partial \xi^+_i}{\partial t} dS_{zi}^+ - \frac{\partial \xi^-_i}{\partial t} dS_{zi}^- \right]}_{\textrm{Extra Time Evolving (TE) Terms} }    
\label{VA_kinetic_correction}
\end{align}
where we have used $\partial \insurfi{$\pm$} / \partial z_i = 0$ to write the expression concisely in vector form. 
We identify the kinetic part with reference to \eq{VA_stress} using an overbrace, under the assumption that everything inside the ${\partial}/{\partial \boldsymbol{r}}$ operator is VA pressure, and identify the two additional terms, $KC$ and $TE$, missing from the VA treatment.

The second term in \eq{mom_timevo} is the configurational part, which for the VA in a moving interface, swapping $\partial \vartheta_{\lambda} / \partial r_{\lambda}$ to $\partial \vartheta_{\lambda} / \partial r$  results in additional terms \citep{Smith_Braga20}, so the full expression with correction term $CC$ is,
\begin{align} 
\displaystyle\sum_{i,j}^N \boldsymbol{f}_{ij} \boldsymbol{r}_{ij} \cdot 
\int_0^1    \frac{\partial \vartheta_\lambda }{\partial \boldsymbol{r}_\lambda}d \lambda
= &-\frac{\partial }{\partial \boldsymbol{r}} \cdot \overbrace{\displaystyle\sum_{i,j}^N \boldsymbol{f}_{ij} \boldsymbol{r}_{ij}  \int_0^1   \vartheta_\lambda d \lambda }^{\PressureVA {\!\!^c} \Delta V  }
\nonumber \\
& + \underbrace{ \displaystyle\sum_{i,j}^N \boldsymbol{f}_{ij} \boldsymbol{r}_{ij} \cdot \int_0^1 \left[\frac{\partial \insurfl{+} }{\partial \boldsymbol{r}_\lambda}  dS_{z\lambda}^+ - \frac{\partial \insurfl{$-$} }{\partial \boldsymbol{r}_\lambda} dS_{z\lambda}^- \right]   d\lambda }_{\textrm{Extra Configurational Curvature (CC) Term} },
\label{VA_config_correction}
\end{align}
again using $\partial \insurfl{$\pm$} / \partial z_\lambda = 0$ to write in vector form. 
So, \eq{mom_timevo} shows the link between SF and VA form with the extra terms naturally included in the SF derivation,
\begin{align} 
\frac{d}{d t} \displaystyle\sum_{i=1}^N  m_i \boldsymbol{\dot{r}}_i  \vartheta_i  =  - \!\!\!\! \sum_{\alpha \in \{x,y,z\}}  \!\! [\StressSF{_{\alpha}^+ } - \StressSF{_{\alpha}^-}  ] \Delta S_{\alpha}  \approx -\frac{\partial }{\partial \boldsymbol{r}} \cdot  \PressureVA \Delta V + KC + TE + CC 
\end{align}
where the time evolution is exactly equal to the sum of SF over all six faces (convective terms are set to zero for simplicity) but only approximately equal to the divergence of the volume average with correction terms \footnote{The exact equality is lost because the VA form assumes pressure is constant in a volume, while the exact balance of the control volume requires fluxes averaged over the bounding surfaces of the volume.}.

Pressure in the VA form is the average inside a divergence operator ${\partial }/{\partial \boldsymbol{r}}$, so it is natural to write the extra terms in Eqs. (\ref{VA_kinetic_correction}) and (\ref{VA_config_correction}) in the same way.
Terms in the $x$ and $y$ directions are expected to be zero so we take the derivative of the integral in just the $z$ direction, 
\begin{align} 
\frac{\partial}{\partial z} \int \left[ KC + TE + CC \right]dz \;\;\;\;\;\;\;\;\;\;\;\;\;\;\;\;\;\;\;\;\;\;\;\;\;\;\;\;\;\;\;\;\;\;\;\;\;\;\;\;\;\;\;\;\;\;\;\;\;\;\;\;\;\;\;\;\;\;\;\;\;\;\;\;\;\;\;\;\;\;\;\;\;\;\;\;\;\;\;\;\;\;\;
\nonumber \\
=  \frac{\partial}{\partial z} \left[ \displaystyle\sum_{i=1}^N  \left( m_i {\dot{z}}_i  \boldsymbol{\dot{r}}_i \cdot \frac{\partial \xi_i}{\partial \boldsymbol{r}_i}     +  m_i {\dot{z}}_i \frac{\partial \xi_i}{\partial t}\right)  \vartheta_i  +  \displaystyle\sum_{i,j}^N {f}_{zij} \boldsymbol{r}_{ij} \cdot \int_0^1 \frac{\partial \insurfl{} }{\partial \boldsymbol{r}_\lambda}  \vartheta_{\lambda} d\lambda  \right]
\label{Extra_Terms_VA}
\end{align}
so everything inside the derivative is in the form of a pressure contribution inside a volume.
As with other VA terms, the functions $\vartheta_{\alpha}$ assigns pressure contributions only when molecules, $\alpha=i$, or points on interactions path, $\alpha=\lambda$, are inside the volume.
The interface movement  ${\partial \xi_i}/{\partial t}$ and surface curvature  ${\partial \xi_i}/{\partial \boldsymbol{r}_i}$ at the location of the molecules is used in the kinetic part, while the curvature integrated along the contour between molecules ${\partial \insurfl{} }/{\partial \boldsymbol{r}_\lambda} $ appears in the configurational part.
Obtaining curvature at varying locations while moving along an interaction contour will likely be complex and computationally expensive.
For simplicity, these terms will be estimated using the surface flux terms from \eq{z_stress} in this work as these are already obtained when collecting SF expressions and are expected to be similar for the thin averaging volumes employed.

\section{Methodology}
\label{sec:methods}

\begin{figure}
  \centering
    \includegraphics[width=1.0\textwidth]{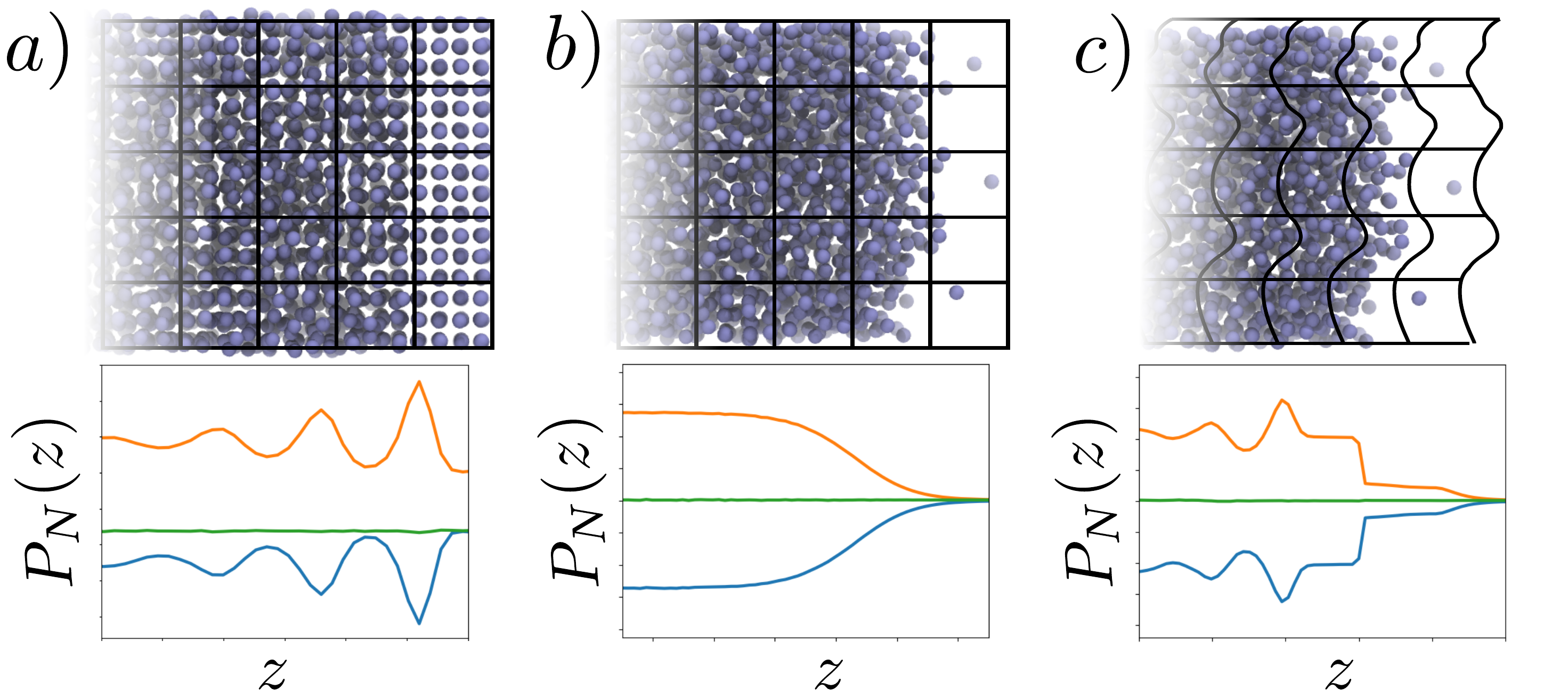}
      \caption{A schematic of the three cases considered in this work $a)$ a solid-liquid interface with a fixed grid, $b)$ a liquid-vapour interface with a fixed grid and $c)$ a liquid-vapour interface with a grid moving with the intrinsic interface. The profiles of normal pressure are show below with kinetic (\textcolor{c2}{\xlinethick}), configurational (\textcolor{c1}{\xlinethick}) and total (\textcolor{c3}{\xlinethick}) pressure contributions.}
      \label{Cases}
\end{figure}

In this section, we outline the molecular dynamics (MD) setup for the three cases shown schematically in Figure \ref{Cases}.
All cases use the simple, pairwise shifted Lennard Jones potential,
\begin{align}
\phi(r_{ij}) = \epsilon \left[ \left(\frac{\sigma}{r_{ij}}\right)^{12} - \left(\frac{\sigma}{r_{ij}}\right)^{6} \right] - \phi(r_c); \;\;\;\;\; r_{ij} < r_c,
\label{Lennard_Jones}
\end{align}
with a cutoff $r_c=2.5$, where all numbers are presented in reduced LJ units.
This is chosen to allow efficient simulations but is shorter than required to give surface tension with good experimental agreement \citep{Smith_et_al16}.
Time integration uses the velocity Verlet scheme,
\begin{align}
\boldsymbol{r}_i(t+\Delta t) & =\boldsymbol{r}_i(t) + \Delta t \boldsymbol{v}_i(t+\Delta t/2) \nonumber \\
\boldsymbol{v}_i(t + \Delta t /2 ) & = \boldsymbol{v}_i(t- \Delta t/2) + \Delta t  \boldsymbol{F}_i(t)
\end{align}
with a timestep $\Delta t = 0.005$ and force on $i$ calculated from $\boldsymbol{F}_i = \sum_{j \ne i}^N\boldsymbol{f}_{ij}$.
The simulations are run using Flowmol which has been verified in a range of previous publications \citep{Smith_Thesis, Smith_2015, Smith_Braga20} and has recently been made open-source \citep{FlowMol}.

For the solid-liquid simulation, the domain is setup with height $Lz=34.2$ and walls of thickness $3$ from the top and bottom of the domain.
The system is periodic in the other directions, with $Lx = Ly = 13.6$ giving a total of $N=5120$ molecules.
The walls are fixed using tethered molecules with anharmonic spring constants of \citet{Petravic_Harrowell} with strength $k_4 =5\times 10^3$ and $k_6 =5\times 10^6$. 
The system is initialised using an FCC lattice of density $\rho = 0.8$ throughout and initial temperature of $T=1$, with the untethered region allowed to melt while the system is equilibrated.
After melting, the average temperature of the system is around $T_{ave}=0.6$.
Statistics are then collected over 100,000 steps.

For the liquid-vapour coexistence, the setup is identical to previous work \citep{Smith_Braga20}, with periodic boundaries in all directions of size $Lx = Ly = 12.7$ and $Lz= 47.62$.
The setup also uses an FCC lattice and deletes molecules, with the middle $40\%$ designated as liquid with a density of $\rho_l = 0.79$ and remaining gas at density $\rho_g = 0.0002$ giving $N=2635$ molecules.
The system is equilibrated for 100,000 timesteps in a Nos\`{e} Hoover NVT ensemble with setpoint $T_s=0.7$.
The main runs are restarted in an NVE ensemble run until well-resolved statistics are obtained.

The intrinsic interface is fitted to the outer molecules of a cluster of liquid molecules, identified by a cluster analysis with Stillinger cutoff length $r_d=1.5$ and requiring more than three neighbours per molecule.
The functional form of surface $\xi(x,y,t)$ is fitted using the intrinsic surface method (ISM) \citep{Chacon2003aa} where the interface is described by an arbitrary number of sines and cosines,
\begin{align}
\xi(x,y,t)=\displaystyle\sum_{\mu=-k_u}^{k_u} \displaystyle\sum_{\nu=-k_u}^{k_u}  a_{\mu \nu}(t) f_\mu (x) f_\nu(y),
\label{Matrix_eqn}
\end{align}
where $f_\mu (x) = \cos(k_x x)$, $f_{-\mu} (x) = \sin(k_x x)$ and $ a_{\mu \nu}$ is the matrix of surface wavenumbers.
The number of wavelengths is calculated from the system size $k_u = nint(\sqrt{Lx Ly}/\lambda_{u})$ with $\lambda_{u}=1$ the minimum wavelength.
This function is fitted by minimising the least-square difference between surface molecules $z_p$ and the intrinsic surface function at these positions $\xi = \xi(x_p, y_p,t)$,
\begin{align}
W = \frac{1}{2}\displaystyle\sum_{p=1}^{N_p} \left[ z_p(t) - \xi(x_p, y_p,t) \right]^2 +  \psi \tilde{A},
\label{Minimise}
\end{align}
with extra constraint $\psi=1\times10^{-8}$ to prevent overfitting by ensure intrinsic area $\tilde{A}$ does not become too large.
This process proceeds in stages, starting with a $3$ by $3$ grid of the most extreme molecules, the fitting is repeated, using new pivots added in batches based on proximity to the current surface, until the density of the surface reaches a value of $n_s = 0.7$.
Once fitted, this surface is then used to determine the grid of cells, or bins, which we use to collect averaged statistics.
This is the Lagrangian reference frame used in to obtain pressure measurements which are always the same distance from the current position of the moving surface.
The domain is split into bins with dimensions $\Delta x = \Delta y = 0.198$ tangential to the surface and $\Delta z = 0.0875$ in the normal direction.
For the SF term, the intrinsic interface is converted to a piecewise set of bilinear squares each of size $\Delta x$ by $\Delta y$ which are used to calculate the interaction with the surface.
This employs an efficient algorithm for ray-tracing to obtain crossings \citep{Ramsey_bilinear_2004} and assigns these to the appropriate cells.
The conservation of every control volume in the domain is tested every timestep, by summing up \eq{z_equation_implementation} for all six surfaces and checking it has a difference of less than $1\times 10^{-9}$ compared to the momentum change inside that volume.
For the VA pressure, a mapping based on molecular positions, or points on the interaction contour between molecules, is applied using the corresponding surface position.
The mapped coordinate can then be efficiently binned using integer division, as if on a uniform grid.
The full details of these algorithms are described extensively in previous work \citep{Smith_Braga20}

\section{Results and Discussion}
\label{sec:results}

\begin{figure}
  \centering
    \includegraphics[width=1.0\textwidth]{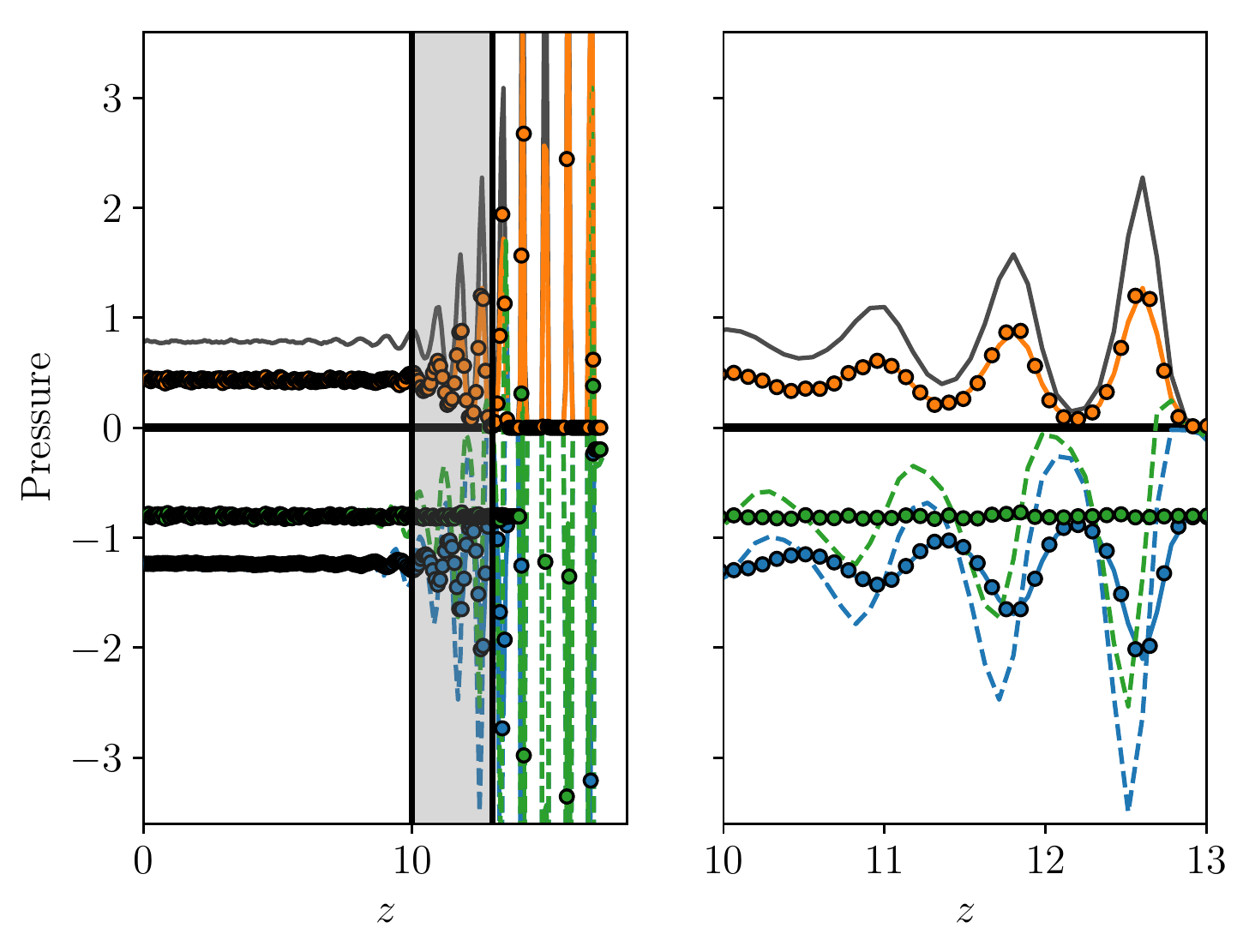}
\put(-350,280){$a)$}
\put(-160,280){$b)$}
      \caption{Comparing wall-normal, $P_{N}$, pressure measurements near a solid-liquid interface showing $a)$ half channel and $b)$ near wall region. The kinetic components are shown for IK1 (\textcolor{c2}{\xdashthick}), VA (\fcirc{c2}) and SF (\textcolor{c2}{\xlinethick}), the configurational part for  IK1 (\textcolor{c1}{\xdashthick}), VA (\fcirc{c1}) and SF (\textcolor{c1}{\xlinethick}) while the total pressure is  IK1 (\textcolor{c3}{\xdashthick}), VA (\fcirc{c3}) and SF (\textcolor{c3}{\xlinethick}). The density (\textcolor{gray}{\xlinethick}) with the zero axis shown by a horizontal black line (\xlinethick) and the shaded region on $a)$ is the section shown in $b)$}
      \label{Solid_liquid_interface}
\end{figure}

\begin{figure}
  \centering
    \includegraphics[width=1.0\textwidth]{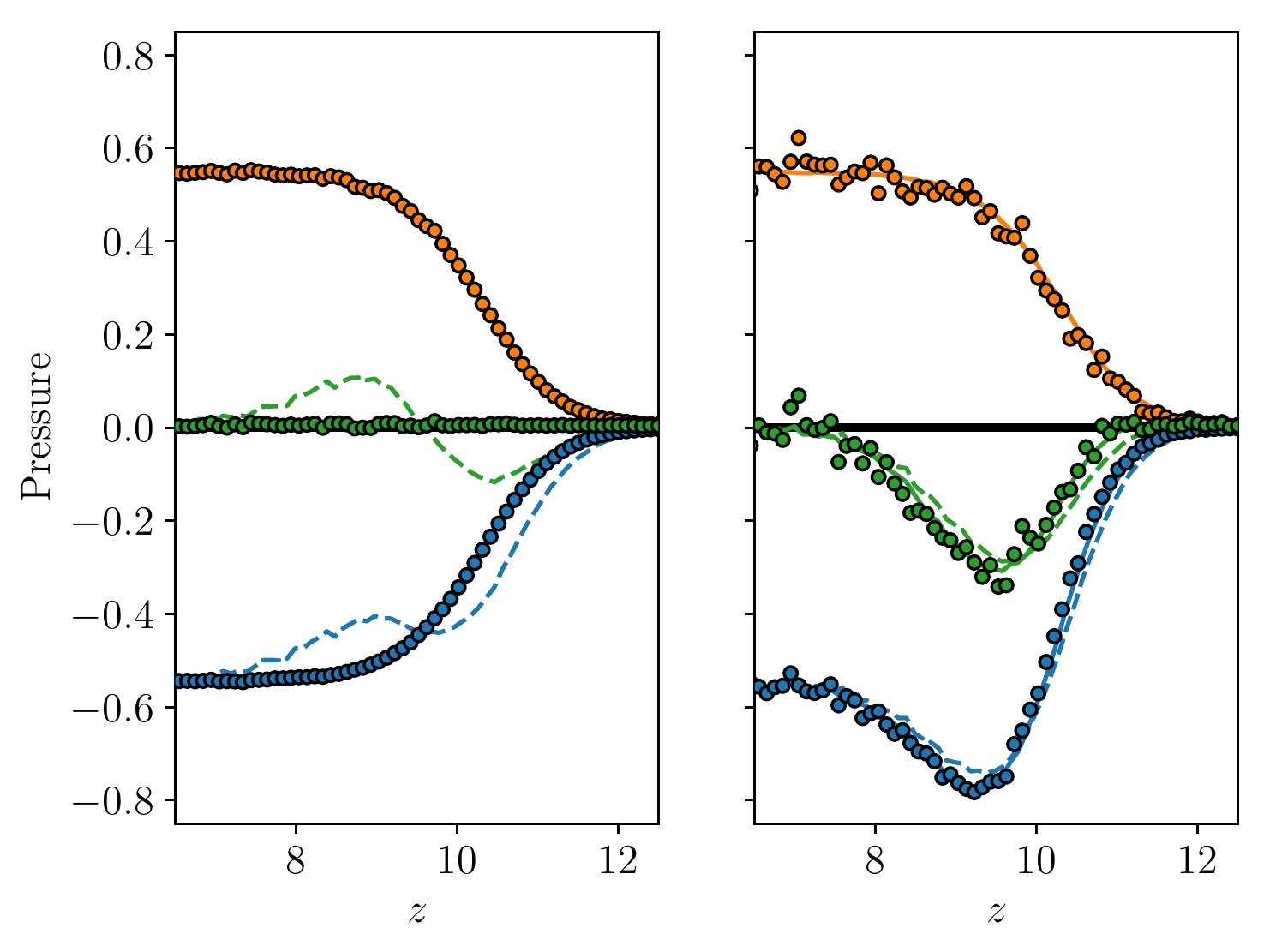}
\put(-346,280){$a)$}
\put(-160,280){$b)$}   
 \caption{$a)$ Normal $P_{N}$ and $b)$ tangential $P_{T}$ pressure near a liquid-vapour interface using a fixed reference frame (a uniform grid). The kinetic components for IK1 (\textcolor{c2}{\xdashthick}), VA (\fcirc{c2}) and SF (\textcolor{c2}{\xlinethick}), the configurational part for IK1 (\textcolor{c1}{\xdashthick}), VA (\fcirc{c1}) and SF (\textcolor{c1}{\xlinethick}) and the total pressure for IK1 (\textcolor{c3}{\xdashthick}), VA (\fcirc{c3}) and SF (\textcolor{c3}{\xlinethick}) with the zero axis shown by a horizontal black line (\xlinethick) . The Irving Kirkwood contour would give the same normal and tangential pressure as the VA and SF curves. }
      \label{interface_fixed_reference}
\end{figure}

\begin{figure}
  \centering
    \includegraphics[width=1.0\textwidth]{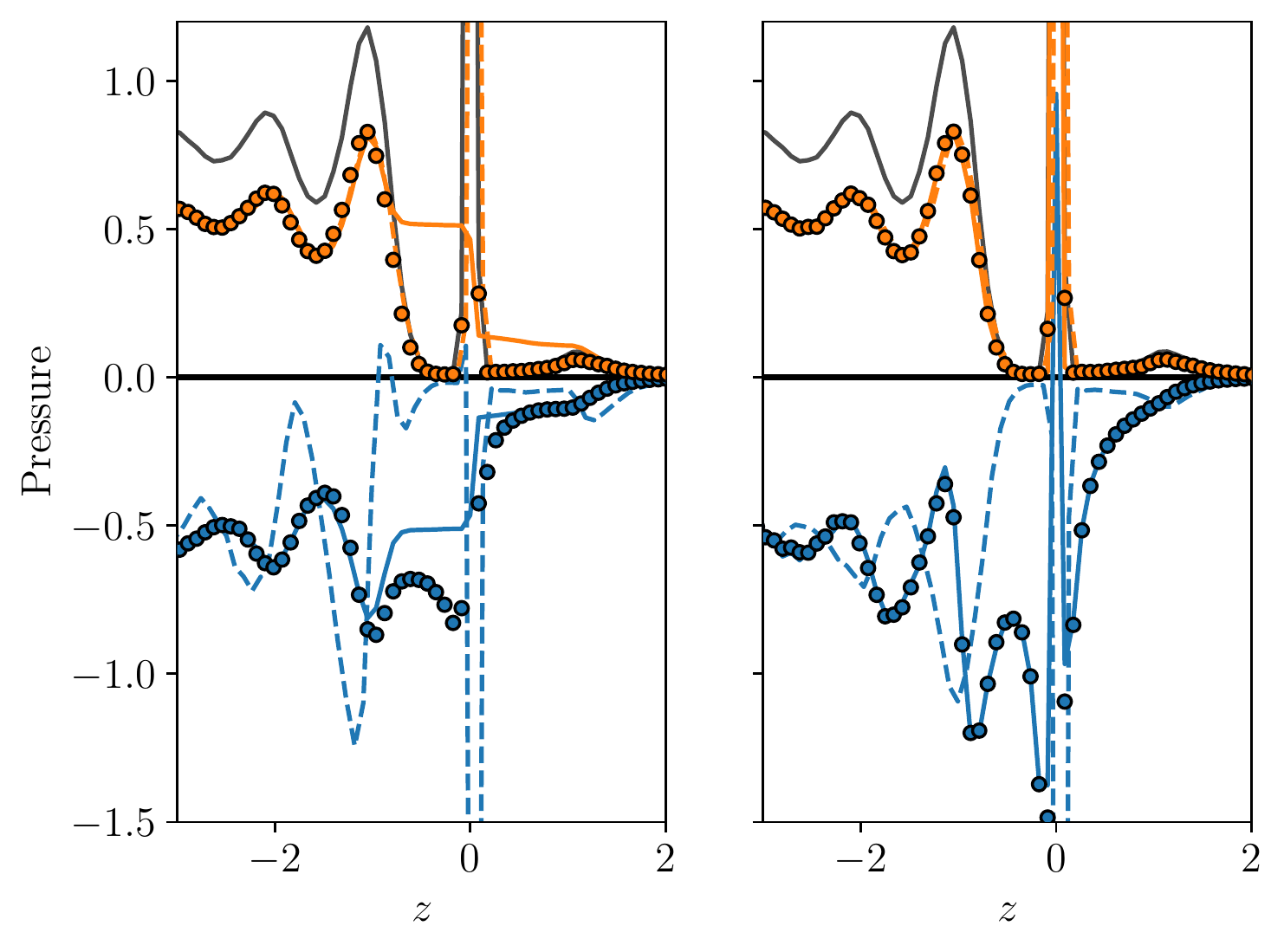}
\put(-346,280){$a)$}
\put(-160,280){$b)$}   
      \caption{ $a)$ Normal $P_{N}$ and $b)$ tangential $P_{T}$ pressure for a reference moving with the liquid-vapour interface, showing the kinetic components for IK1 (\textcolor{c2}{\xdashthick}), VA (\fcirc{c2}) and SF (\textcolor{c2}{\xlinethick}) (note the dashed line exactly follows the solid line) and the configurational part for IK1 (\textcolor{c1}{\xdashthick}), VA (\fcirc{c1}) and SF (\textcolor{c1}{\xlinethick}) with density (\textcolor{gray}{\xlinethick}) with the zero axis shown by a horizontal black line (\xlinethick). The Harasima contour would give the same normal pressure as the SF curves in $a)$ and same tangential pressure as the IK1 in $b)$.}
      \label{interface_moving_reference}
\end{figure}

\begin{figure}
  \centering
    \includegraphics[width=1.0\textwidth]{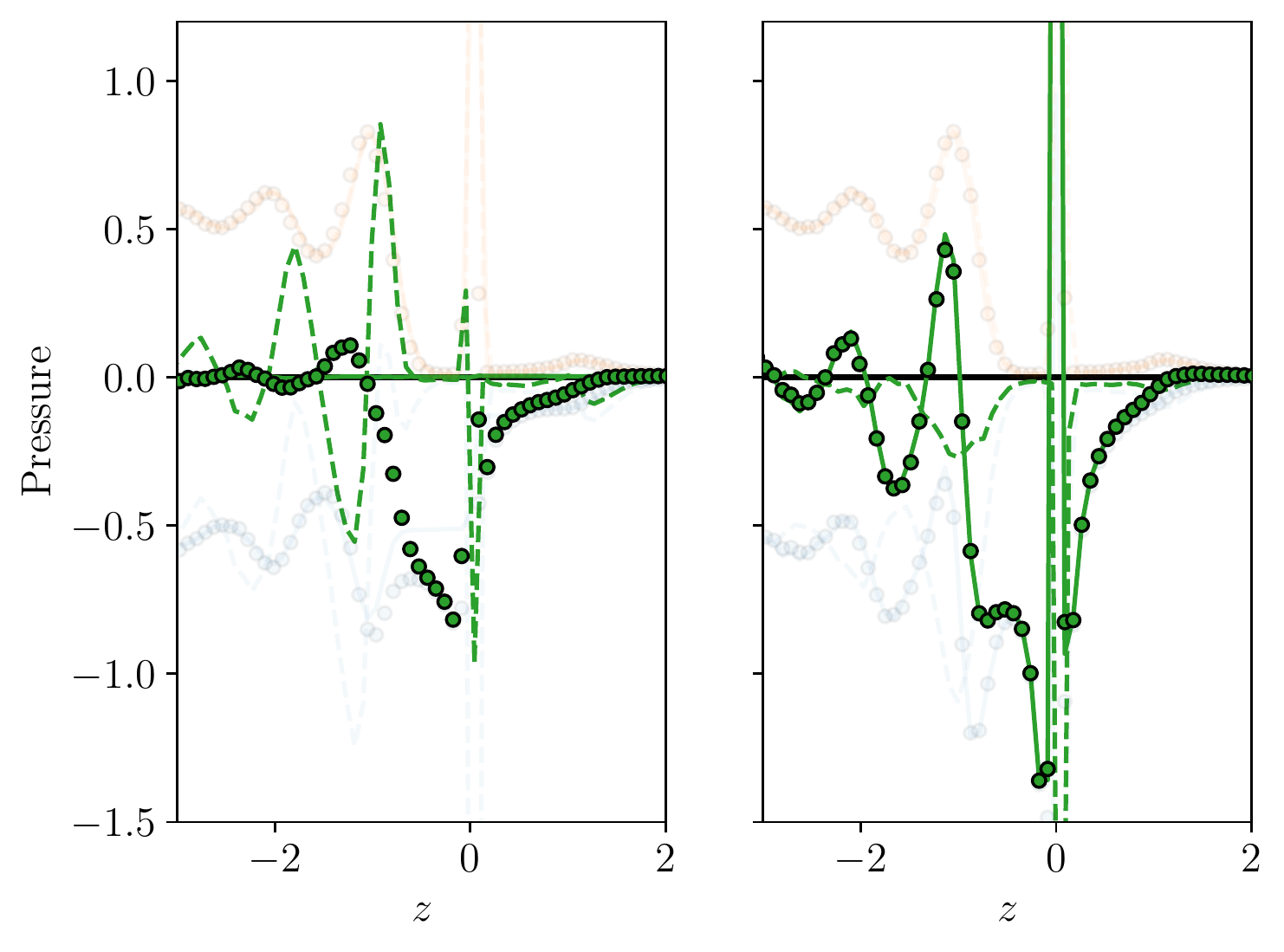}
\put(-346,280){$a)$}
\put(-160,280){$b)$}    
      \caption{The total pressure components for a reference moving with the interface, $a)$ Normal $P_{N}$ and $b)$ tangential $P_{T}$,  with IK1 (\textcolor{c3}{\xdashthick}), VA (\fcirc{c3}) and SF (\textcolor{c3}{\xlinethick}) with the kinetic and configurational constituent parts from Fig \ref{interface_moving_reference} shown faintly for reference and zero axis shown by a horizontal black line (\xlinethick) . The Harasima contour would give the same normal pressure as the SF curves in $a)$ and same tangential pressure as the IK1 in $b)$.}
      \label{interface_moving_reference_total}
\end{figure}

\begin{figure}
  \centering
    \includegraphics[width=1.0\textwidth]{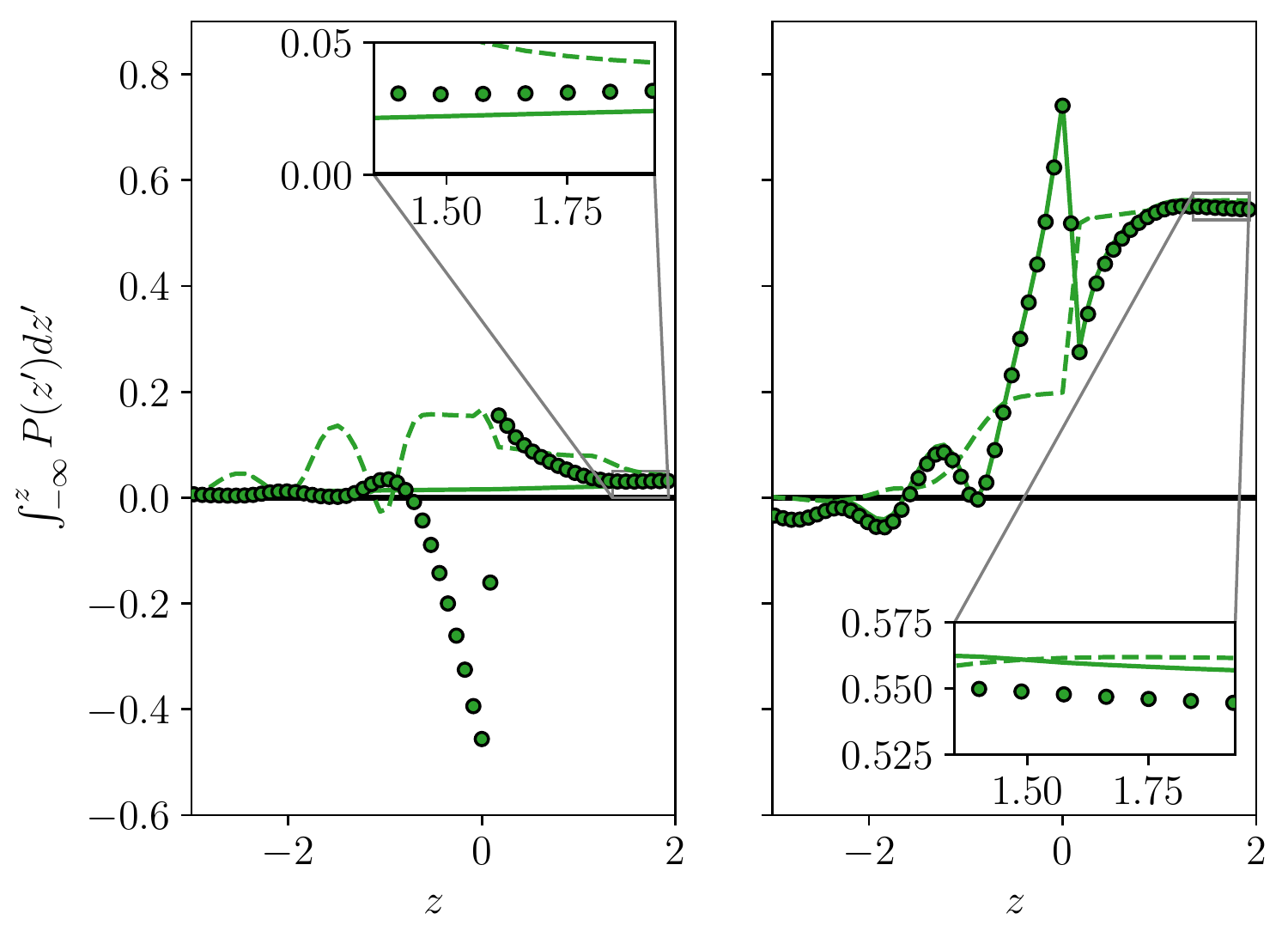}
\put(-340,280){$a)$}
\put(-155,280){$b)$}   
      \caption{The cumulative integral of the total pressures in a reference moving with the interface shown in Figure \ref{interface_moving_reference_total} with normal $P_N$ on $a)$ and the negative of the tangential $-P_T$ pressure shown on $b)$, starting from $z=-5$ up to the current $z$ on the axis, for the IK1 (\textcolor{c3}{\xdashthick}), VA (\fcirc{c3}) and SF (\textcolor{c3}{\xlinethick}) measurements, and the horizontal black line (\xlinevthick) is the zero axis. The inserts zoom in on the values that the integrals have converged to by the upper limit displayed on the plot, $z=2$, where the sum of normal and tangential components would gives the total surface tension.}
      \label{Integral_ST}
\end{figure}

\begin{figure}
  \centering
    \includegraphics[width=1.0\textwidth]{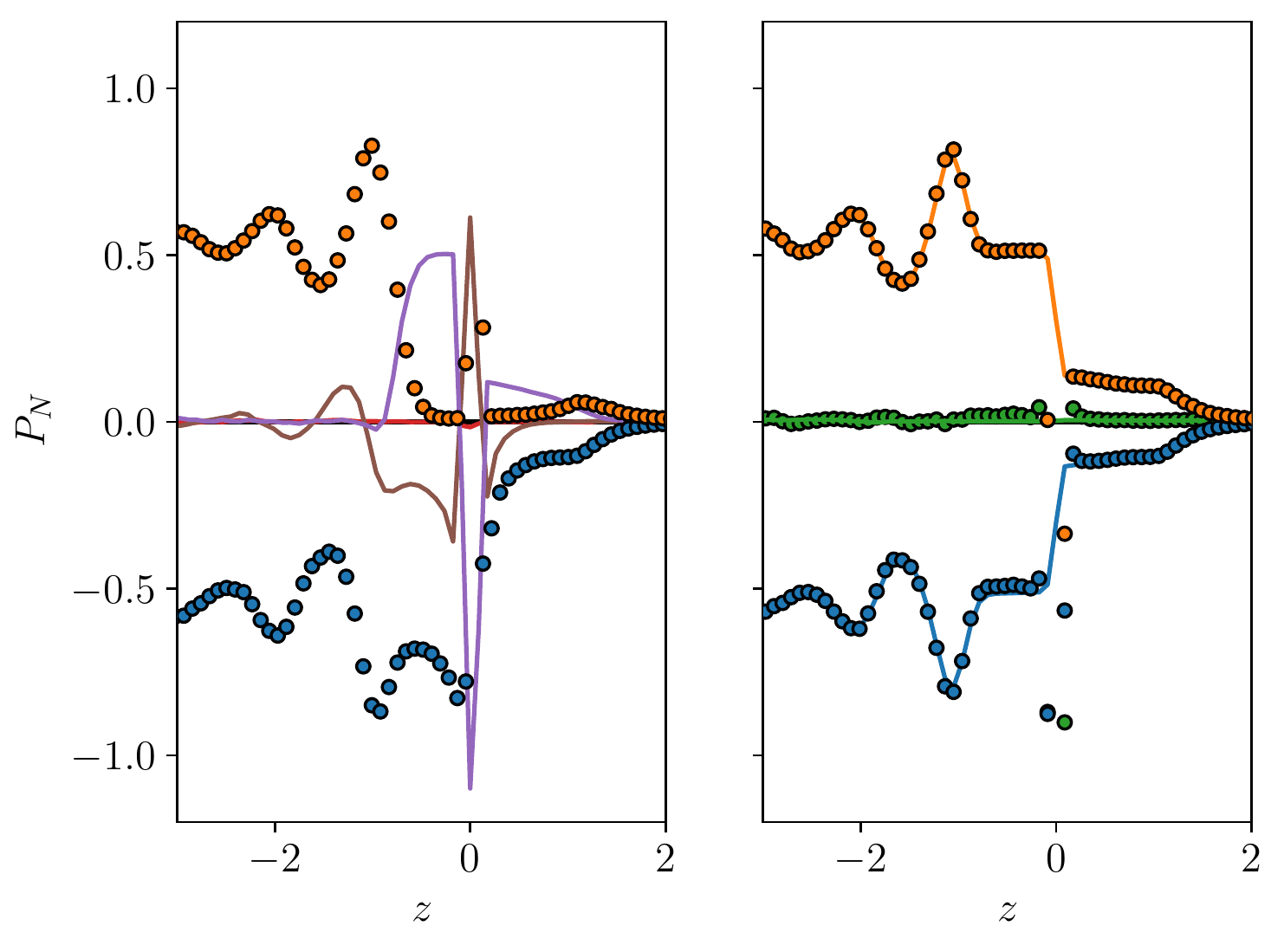}
	\put(-346,280){$a)$}
	\put(-160,280){$b)$}    
      \caption{The terms required to balance normal pressure in a reference moving with the interface, where $a)$ shows the normal component of the Volume Average (VA) pressure, both kinetic (\fcirc{c2}) and configurational (\fcirc{c1}) with additional correction terms from Eqs. (\ref{VA_kinetic_correction}) and (\ref{VA_config_correction}) including kinetic curvature $\partial \xi_i / \partial \boldsymbol{r}_i$, KC, (\textcolor{c4}{\xlinethick}), time evolving $\partial \xi_i / \partial t$, TE,  (\textcolor{c5}{\xlinethick}) as well as the negative of the configurational curvature $\partial \xi_{\lambda} / \partial \boldsymbol{r}_{\lambda}$, CC, (\textcolor{c6}{\xlinethick}) (shown as negative to allow comparison with the VA configurational pressure). In figure $b)$, the VA pressure is displayed with the correction terms added compared to the SF forms, which naturally include all of these terms. These include the kinetic VA with KE and TE added (\fcirc{c2}), kinetic SF (\textcolor{c2}{\xlinethick}), configurational VA with CC added (\fcirc{c1}) against configurational SF (\textcolor{c1}{\xlinethick}) and total corrected VA  (\fcirc{c3}) against total SF (\textcolor{c3}{\xlinethick}). The zero axis is shown by a horizontal black line (\xlinethick).}
\label{CV_terms}
\end{figure}

\begin{figure}
  \centering
    \includegraphics[width=1.0\textwidth]{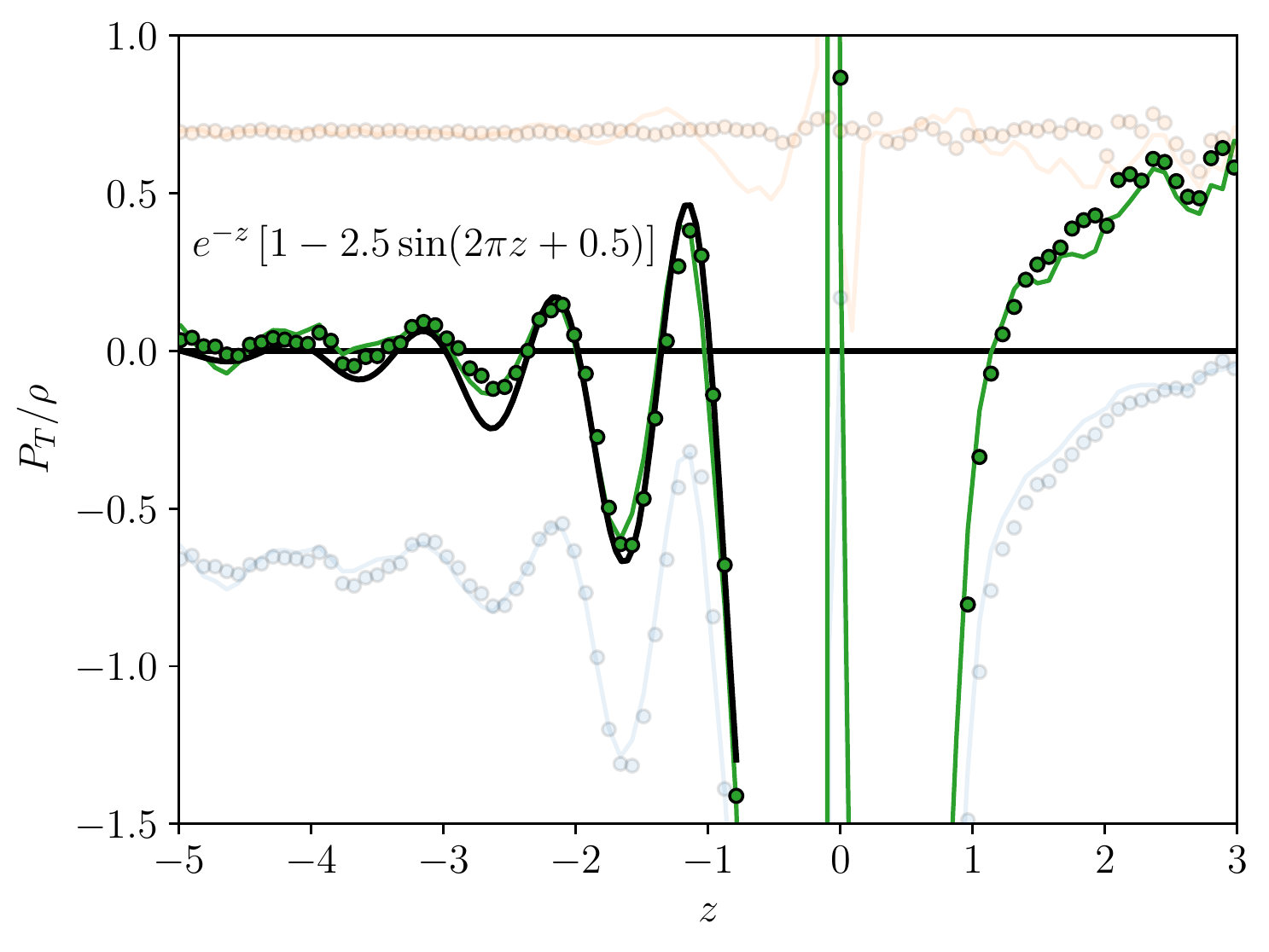}
      \caption{Total tangential pressure divided by density for quantities in a reference frame moving with the interface, VA  (\fcirc{c3}) and total SF (\textcolor{c3}{\xlinethick}). The kinetic, SF (\textcolor{c2}{\xlinethick}), VA  (\fcirc{c2}) , and configurational, SF (\textcolor{c1}{\xlinethick}), VA  (\fcirc{c1}) components are shown faintly in the background for reference. A fitting is shown (\xlinethick) with functional form and fitting coefficients annotated on the figure with the zero axis the horizontal black line (\xlinethick) .}
      \label{pressure_density}
\end{figure}

In this section, we compare three types of pressure measurements, the IK1 \eq{IK1}, VA \eq{VA_stress} and SF \eq{z_stress}, split into kinetic and configurational components.
These three forms also provide the Irving Kirkwood and Harasima contours, and the link to these is discussed.
The convention is the same in all figures, kinetic terms are shown in orange, configurational in blue and total pressures in green. The IK1 pressure is shown as a dotted line, the VA as filled circles and the SF as a solid line.
The three cases are shown in Figure \ref{Cases}, first \ref{Cases}$a)$ the solid-liquid case is shown in Figure \ref{Solid_liquid_interface}, which is used to demonstrate the flat normal pressure profile obtained by both the VA and SF methods, as required for mechanical equilibrium, together with the varying IK1.
The same set of pressure measurements of case  \ref{Cases}$b)$ the liquid-vapour interface using a fixed averaging grid are shown in Figure \ref{interface_fixed_reference}, demonstrating the same flat normal pressure profile  from both VA and SF terms and the same variation in the IK1 expression.
Finally, \ref{Cases}$c)$ is the focus of the remaining work, using interface tracking to obtain pressure in a reference moving with an intrinsic liquid-vapour interface.
All of the IK1, VA and SF forms of pressure give different behaviour, and the kinetic and configurational terms are compared in Figure \ref{interface_moving_reference}, before discussing the total pressure in Figure \ref{interface_moving_reference_total}.
In order to understand the differences, the normal and tangential parts are presented individually and the resulting contributions to surface tension from their respective integrals are displayed in Figure \ref{Integral_ST}.
A full breakdown of all SF contributions to normal pressure from \eq{z_stress}, together with the surface curvature and movement which are needed as corrections to the VA form, then given in Figure \ref{CV_terms}.
This is shown to satisfy the momentum equilibrium in a moving control volume, proving the SF form must contain all possible contributions and that the curvature and surface movement are essential to ensure VA expressions satisfied mechanical equilibrium in the reference frame moving with the interface.
The shape of tangential pressure is shown normalised by density in Figure \ref{pressure_density} with a fitting proposed which could be used to approximate pair interactions in terms of density.
The link to the Irving Kirkwood and Harasima contours is discussed.

We start with the case shown in the schematic of Figure \ref{Cases} $a)$.
The wall normal ($P_N = P_{zz}$) pressure near a solid-liquid interface is presented first in Figure \ref{Solid_liquid_interface} with half of the channel shown in $a)$ and a focus on just the near-wall region in $b)$.
The shaded region in Figure \ref{Solid_liquid_interface}$a)$ shows the area focused on in \ref{Solid_liquid_interface} $b)$, as it contains the majority of the variation, and in all subsequent figures only a small region near the interface is displayed.
The kinetic and configurational parts of the pressure are shown to balance for both the VA and SF forms of pressure, with the average giving a constant pressure value. 
This satisfies the condition for equilibrium $\boldsymbol{\nabla} \cdot \boldsymbol{P} = 0$.
The IK1 pressure has identical kinetic component to both the VA and SF, but the configurational part is seen to have larger shifted peaks which results in a non-zero average pressure so $\boldsymbol{\nabla} \cdot \boldsymbol{P} \ne 0$.
These peaks in the pressure cannot be correct in an equilibrium system, as they would induce a flow, so this suggests the IK1 pressure is not correct, a well documented result in the literature \citep{Todd_et_al_95, Varnik_et_al(2000)}.

The same quantities discussed for the solid-liquid case are shown for the liquid-vapour interface using a fixed grid in Figure \ref{interface_fixed_reference}, the schematic of Figure \ref{Cases} $b)$.
As with the solid-liquid cases, the VA and SF gives similar results while the IK1 shows a difference in the normal configurational pressure, the blue line in Figure \ref{interface_fixed_reference} $a)$.
The result again shows a flat line for the VA and SF forms, indicating the required $\boldsymbol{\nabla} \cdot \boldsymbol{P} = 0$ condition is satisfied.
The normal component of SF pressure, shown by the solid lines, is equivalent to the Irving Kirkwood contour. 
The IK1 fails to give the correct normal component, although the difference can be seen to have roughly equal positive and negative areas.
This means the \citet{Kirkwood_Buff} integral used to get surface tension will be the same as the other methods.

The tangential pressure \ref{interface_fixed_reference} $b)$ is almost identical for IK1, VA and SF measurements.
The IK1 shows a slight shift toward the interface when compared to the other forms of pressure.
We do not explicitly calculate the tangential Irving Kirkwood contour of \eq{IK_Tangential}, but it is a combination of the VA form measured on a SF style $z$ plane, so we would expect it to give the same results as both the VA and SF expressions, which are very similar in Figure \ref{interface_fixed_reference} $b)$.

So, the conclusion from Figures \ref{Solid_liquid_interface} and \ref{interface_fixed_reference} are that the VA and SF satisfy force balance in the normal direction while the IK1 does not.
However, the IK1 $P_N$ contribution above and below the axis are roughly equal so this will cancel in an integral.
As the tangential components of the IK1 is similar to both the VA and SF, the integral of $P_N - P_T$ used in surface tension will be identical.
For the normal pressure, the Irving Kirkwood contour is represented by the SF results and satisfy the equilibrium $\boldsymbol{\nabla} \cdot \boldsymbol{P} = 0$.
For the tangential pressure, the Irving Kirkwood contour is a combination of the VA and SF, which give very similar results.
There is very little difference in these various measures for a fixed reference frame, so next we consider the same pressure measurements in a reference frame moving with the interface, the case shown in Figure \ref{Cases} $c)$. 

Both kinetic and configurational components of pressure are shown in Figure \ref{interface_moving_reference} with the total pressure shown in Figure \ref{interface_moving_reference_total}, again split into normal $a)$ and tangential components $b)$.
As before, the IK1, VA and SF terms are compared, but here we see notable differences in all three.
For the kinetic terms in Figure \ref{interface_moving_reference}$a)$, the IK1 and VA agree, showing peaks which mirror the density shown in gray, representing the location of the molecules relative to the surface. 
The SF term is shown and naturally includes the surface evolution term, $\partial \xi / \partial t$ and kinetic curvature which are not included in the VA expression.
The kinetic curvature terms are zero on average, but the inclusion of the surface evolution is important, without it the kinetic SF pressure profile would be identical to the VA and IK1 \citep{Smith_Braga20}.
As a result of including the surface movement, the SF profile has a flat region on the liquid side (orange line Figure \ref{interface_moving_reference}$a)$), between the interface and first layer, as well as a smoother transition to zero on the gas side.
The profile is exactly mirrored by the SF configurational pressure, the solid blue line on \ref{interface_moving_reference}$a)$, so the sum of kinetic and configurational terms gives a flat pressure profile on Figure \ref{interface_moving_reference_total}$a)$, the solid green line.
We therefore see from Figure \ref{interface_moving_reference_total}$a)$, only the SF form satisfies the equilibrium force-balance condition. 
The extra term in \eq{VA_config_correction} for the configurational curvature is also missing from the configurational term of the VA pressure, a point we return to at the end of this section.
As a result, the VA pressure does not satisfy the equilibrium force-balance condition in \ref{interface_moving_reference_total}$a)$ in a moving reference frame.
The IK1 form is very different to the VA and SF, showing large oscillations in normal configurational pressure which results in significant peaks in the configurational pressure of Figures \ref{interface_moving_reference}$a)$ and therefore the total in \ref{interface_moving_reference_total}$a)$.
As the IK1 pressure fails to satisfy the balance condition in the simpler case of a solid-liquid interface, as well as the fixed reference liquid-vapour case, there is no reason to suspect it will perform any better for a moving-reference case.
In fact, the oscillatory nature of the IK1 pressure appears to show more prominent peaks which is a feature of the IK1 observed in the solid-liquid case of Figure \ref{Solid_liquid_interface}.
This is interesting as the IK1 departs further as the system becomes more inhomogeneous and structured \citep{Varnik_et_al(2000)}, so the IK1 performs worse here as a moving reference frame exposes the inhomogeneity of the molecular structure near the intrinsic interface.
These molecular peaks are apparently smoothed by the continuous operation of taking a fraction of line inside a volume for the VA case, while the SF expression depends on crossings and so is less dependent on the absolute position of the molecules.
The measuring volumes are also centred on the interface itself so surfaces crossing measurements are offset either side by $\Delta z/2$.
The SF therefore avoids the peak at the interface in the normal pressure measurement seen in Figure \ref{interface_moving_reference} $a)$.

The tangential components are identical for VA and SF in \ref{interface_moving_reference_total}$b)$, as they are both calculated using a uniform grid in the surface tangent direction, so we would expect them to agree in the same way any fixed volume cases do.
The IK1 actually shows smaller oscillation than the VA/SF terms in the tangential direction with a peak at the interface and a smaller, more smeared contribution for the first and second fluid layers.

%

Despite the very different profiles for the three pressure measurements, it is well-documented that the \citet{Kirkwood_Buff} integral will give the same surface tension for all three \citep{Malijevsk__2012}.
This integral of the difference between normal and tangential terms will also be the same as using a fixed reference frame \citep{Smith_Braga20}.
One way to understand this: all definitions of pressure considered here use the same forces but vary how they distribute these measurements to different bins or cells, so the integral must be the same.
As the \citet{Kirkwood_Buff} integral is using the difference between normal and tangential components $\int (P_N - P_T) dz$, it is instructive to split them into these two additive contributions, as is done in Figure \ref{Integral_ST} with $a)$ the integral of the normal $\int P_N  dz$ and $b)$ the integral of the negative of the tangential $-\int  P_T dz$.
In Figure \ref{Integral_ST} $a)$, the flat SF normal pressure profile gives a linearly increasing integral over the interface which agrees with the oscillating VA and IK1 after the integral has moved far enough away from the interface into the gas phase (shown approaching $z=2$ in the insert).
The integral of the normal contribution is small, contributing less than 10\% ($0.05$ by $z=2$) to the total surface tension integral which is approximately $0.6$ in this case). 
The main contribution is from the tangential pressure which shows an identical trend for both the VA and SF in Figure \ref{Integral_ST} $b)$.
The tangential part of the IK1 pressure can be seen to have fewer fluctuations and therefore a simpler integrated contribution.
Again, all three methods converge to approximately the same integrated result for the tangential contributions, with a value of about $0.55$ by $z=2$.
Therefore, the measured surface tension, obtained from the sum of the normal, \ref{Integral_ST}$a)$ and minus tangential \ref{Integral_ST}$b)$, converges to roughly the same value of $\gamma = \int  P_N - P_T dz  \approx 0.6$ for all methods.
Differences are attributed to statistical noise, which integrated quantities like surface tension are more susceptible to.

Figure \ref{CV_terms}$a)$ shows the VA pressure and the three additional terms, KE, TE and CC, which were derived as corrections to the VA pressure in section \ref{sec:VA_correction}.
These terms are due to the moving interface and would be zero for a fixed reference frame, which is why the VA and SF normal pressures have identical profiles in Figures \ref{Solid_liquid_interface} and \ref{interface_fixed_reference} but not in Figures \ref{interface_moving_reference} and \ref{interface_moving_reference_total}.
The corrected VA forms are shown to give a constant normal pressure in Figure \ref{CV_terms}$b)$, as required for mechanical equilibrium in a moving interface, as well as good agreement with the SF pressure which naturally include these terms.
For the kinetic pressure, the normal VA contribution is shown by orange circles on Figure \ref{CV_terms}$a)$ and the additional correction terms due to kinetic and configurational interface curvature as well as the movement of the surface itself, are shown with lines because they are obtained from SF terms.
The VA shown by orange circles on Figure \ref{CV_terms}$b)$ includes the surface movement correction, the purple line on \ref{CV_terms}$a)$, and the kinetic curvature which is the red line (both extra terms from \eq{VA_kinetic_correction}). 
With these additional corrections, the kinetic part of the VA has identical shape to the SF pressure which naturally includes these terms.
For the configurational pressure, the brown line on \ref{CV_terms}$a)$ shows the magnitude of the configurational curvature contributions, which are the extra terms missing from the VA measurements in \eq{VA_config_correction}.
The total for both SF and corrected VA is shown in green as a flat normal pressure which is indicative of mechanical equilibrium.
However, the VA shows a few values which are not flat near the interface, attributed to using the SF form of pressure to get the correction terms instead of the VA form. 
This was expected to gives a small error for the thin bins used here, but near the rapid change at the interface the limits of this assumption appear to show.
This indicates how important getting an exactly a consistent form of pressure is, with only the SF able to give a mechanical balance at all points.

Finally the tangential pressure is considered in Figure \ref{pressure_density}, where the value is plotted normalised by density.
The peaks in the kinetic pressure $P_T(z)$ and density $\rho(z)$ correspond exactly, as a function of $z$, so the ratio is constant giving a flat profile
The shape of the total pressure is therefore due to the configurational part, which depends on interactions passing through a location in space, and will be non-zero even where no molecules are located.
As a result, a diverging profile is observed either side of zero, as the intrinsic fitting process results in a large density peak at the interface and no molecules either side.
This mean $P_T/\rho \to \infty$ as $\rho \to 0$ but a finite value is observed at zero due to the molecules which sit on the intrinsic surface.
The shape of the profile represent the pressure per molecule, so it is useful to understand the functional form it takes.
Fitting is obtained using a decaying exponential $e^{-a z}$ times an oscillating function $1-b\sin(2 c \pi z + d)$, where $a, b, c \text{ and } d$ are coefficients to be fitted, where a reasonable fit is observed for values $a=1$, $b=2.5$, $c=1$ and $d=0.5$, so tangential pressure in the liquid can be approximated using,
\begin{align}
P_T(z) \approx \rho(z) e^{-z} \left[ 1 - 2.5 \sin \left(2 \pi z + 0.5 \right) \right] \text{   where    } -\infty < z < -1.
\end{align}
This form seems to be a reasonable approximation for the first two peaks and captures the decay to zero moving into the bulk.
This fitted expression could be used in a course-grained approximation or mean-field approach to model the pressure near the interface.

Finally we consider the link between results obtained here in a reference frame tracking the interface and the contour forms.
The Irving Kirkwood contour assumes a fixed reference, so is not trivially related to the forms of pressure shown in Figure \ref{interface_moving_reference} and \ref{interface_moving_reference_total} for a moving reference frame.
However, the Harasima contour is obtained by moving first tangentially to the surface, then in the normal direction.
The tangential part from \eq{VA_Harasima_stress} is therefore a similar quantity to the tangential IK1 components shown in Figures \ref{interface_moving_reference}$b)$ and \ref{interface_moving_reference_total}$b)$.
This shows much smaller oscillations in \ref{interface_moving_reference_total}$b)$ than the VA and SF methods.
The normal component of the Harasima contour is mathematically similar to the SF form, which takes the contribution dotted with the surface normal.
It can be seen in Figure \ref{Integral_ST} that the integral of the normal and tangential terms all converge to the same value, so mixing the tangential components of the IK1 pressure with the normal components of the SF to obtain the Harasima contour would result in the same surface tension.
The Harasima contour would therefore satisfy the equilibrium condition in the normal direction, give the simple tangential profile of  \ref{interface_moving_reference_total}$b)$ and return the correct surface tension.
This work therefore gives the mathematical process to measure this contour for a general surface, by taking the SF in the normal direction \eq{z_equation_implementation} and then the IK1 for the tangential components using \eq{VA_Harasima_stress}.
This process was first reported in the work of \citet{Sega_et_al15} with differences between the Harasima and IK tangential pressure explored in later work \citep{Sega_et_al16}

However, for a general surface, the split of the Harasima contour into a normal and tangential part is non-unique, as highlighted in Figure \ref{schematic_surface} where different combinations of tangent and normal to the surface would be possible.
As calculation of the SF interaction and surface normal are required to get the Harasima contour for a general interface, simply using the SF is preferable in this case, although for more complex potentials and long-range interaction the Harasima contour may offer additional advantages \citep{Sega_et_al16, Shi_et_al20}.
The SF provides three components per surface, so all nine pressure terms are available, also giving the curvature contributions and surface movement shown in Figure \ref{CV_terms}.  
It also provides an explicit mathematical link between any contour crossing the closed surface and momentum evolution inside that closed control volume. 
By mixing the tangential IK1 and SF, we lose this property of exact control volume balance and the guarantee all pressure components are captured.
However, using the SF form moving with the interface provides a pressure which naturally includes all terms and satisfies the required condition of mechanical equilibrium.

%

\section{Conclusions}
\label{sec:conclusions}

The appropriate definition of pressure in molecular dynamics (MD) simulation has been the subject of some debate.
In this work, it is argued that a consideration of the reference frame is more illuminating in defining pressure than the interaction contour.
This is especially true near an interface.
By considering a reference frame described by an arbitrary function $\xi = \xi(x,y,t)$ fitted to the surface every time, we get a general form of interface pressure.
The derivation provides an expression in terms of the surface fluxes (SF), a generalisation of the method of planes (MOP) pressure to an arbitrary surface.
This is compared to a range of other pressure definitions, with the differences discussed.
These include a contour between two molecules averaged in a volume called the volume average (VA) pressure, the virial applied locally with interaction split into two called the IK1 pressure as well as the Irving Kirkwood and Harasima contours.
All forms of pressure give the same surface tension but result in different pressure distributions over an interface.
Results show the IK1 form does not satisfy mechanical equilibrium in the interface normal direction, even in the simplest cases, while the VA fails to  satisfy mechanical equilibrium in the case moving with a reference frame.
Correction terms are derived for the VA form and shown to account for moving interface and curvature, giving the expected results.
Combining the SF and IK1 results is shown to provide a generalisation of the Harasima contour for an arbitrary surface.
As a result, we show choice of reference frame provides a clear link between the various forms of pressure.
The SF is argued to be the preferred form as it naturally includes all curvature and surface movement terms, provides an exact link between SF pressure and instantaneous momentum change in the volume and as six SF pressures bound a closed volume, it ensures any of the infinite possible contour choices must cross one of the surfaces and be included in the pressure tensor.

%
%
%

%
%
%
%

\section*{Acknowledgements}

The author is grateful to the UK Materials and Molecular Modelling Hub for computational resources, which is partially funded by EPSRC (EP/P020194/1 and EP/T022213/1)
%
%

\end{document}